\documentclass[10pt,aps,twocolumn,floats,floatfix,amssymb,prd,superscriptaddress,nofootinbib]{revtex4-2}

\usepackage{graphicx}
\usepackage{amsmath,amssymb}
\usepackage{amsfonts}
\usepackage{xspace} 
\usepackage[usenames]{color}
\usepackage{dcolumn}
\usepackage{bm}
\usepackage{mathrsfs}
\usepackage[colorlinks=true]{hyperref}
\usepackage[all]{hypcap} 
\usepackage[utf8]{inputenc} 
\usepackage{multirow}
\usepackage{etoolbox}
\usepackage{tikz}
\usepackage[normalem]{ulem}
\usepackage{yfonts}

\newcommand{\dd}{\mathrm{d}}

\begin{document}

\title{Inflation, black holes with primary hair, and regular planar black holes from an infinite tower of regularized Lovelock-Proca corrections}


\author{Pedro G. S. Fernandes}
\email{fernandes@thphys.uni-heidelberg.de}
\affiliation{Institut f\"ur Theoretische Physik, Universit\"at Heidelberg, Philosophenweg 12 \& 16, 69120 Heidelberg, Germany}

\author{Jingqian Gou}
\email{gou@thphys.uni-heidelberg.de}
\affiliation{Institut f\"ur Theoretische Physik, Universit\"at Heidelberg, Philosophenweg 12 \& 16, 69120 Heidelberg, Germany}

\author{Lavinia Heisenberg}
\email{L.Heisenberg@thphys.uni-heidelberg.de}
\affiliation{Institut f\"ur Theoretische Physik, Universit\"at Heidelberg, Philosophenweg 12 \& 16, 69120 Heidelberg, Germany}

\author{Nadine Nussbaumer}
\email{nussbaumer\_n@thphys.uni-heidelberg.de}
\affiliation{Institut f\"ur Theoretische Physik, Universit\"at Heidelberg, Philosophenweg 12 \& 16, 69120 Heidelberg, Germany}

\begin{abstract}
Infinite towers of higher-order corrections to General Relativity have been proposed as a mechanism to resolve singularities in early-universe cosmology and black holes, in a variety of settings. In this work, we consider an infinite tower of higher-order Proca corrections inspired by dimensional regularizations of Lovelock invariants. We find that the Big Bang singularity present in General Relativity is replaced by an inflationary epoch. Furthermore, the Lovelock-Proca tower allows for regular planar black hole solutions and spherically symmetric black holes with primary hair.
\end{abstract}

\maketitle

\section{Introduction}
General Relativity (GR) stands as the most successful and extensively tested theory of gravity. However, it faces significant observational and conceptual challenges: Notably, it cannot explain dark matter, which is required to account for phenomena such as galactic rotation curves, gravitational lensing, and large-scale structure formation. Additionally, mounting evidence suggests that dark energy, the driver of the universe's accelerated expansion, is not simply a cosmological constant \cite{DESI:2024mwx,DES:2024jxu,Brout:2022vxf,Rubin:2023jdq}.

In addition, the large-scale correlations of matter and radiation throughout the observable universe indicate the presence of nearly scale-invariant primordial fluctuations, remnants of the universe's earliest moments, which is an epoch that remains largely mysterious. The spectral features of these fluctuations align with what would be expected from nearly-Gaussian quantum fluctuations that were enormously stretched during a brief phase of rapid accelerated expansion, known as \emph{inflation} \cite{PhysRevD.23.347,Starobinsky:1980te,Sato:1981qmu,Linde:1981mu,Albrecht:1982wi,Mukhanov:1990me,Starobinsky:1979ty}, which cannot be explained by GR without extra assumptions. While there is certainly no shortage of inflationary models, ranging from the heuristic scalar field in a slow-roll potential to effective field theory (EFT) approaches for single and multiple fields, the exact nature of inflation remains unknown due to the lack of observational data beyond the Gaussian statistics.

Aside from these cosmological arguments, a fundamental shortcoming of GR lies in its prediction of singularities, both at the very beginning of the universe as well as within black hole interiors. Singularities pose a significant threat to the validity of physical theories, as they mark regions where the theories themselves break down and hence all predictive power is lost. In GR, singularities cause spacetime to be geodesically incomplete and physical observables, such as tidal forces, to diverge.
All of these issues, and a few others, such as the $H_0$ and $\sigma_8$ tensions, provide compelling evidence for new physics in gravity \cite{DiValentino:2021izs,Chen:2024vuf}.

However, going beyond the GR paradigm is not straightforward: as Lovelock's theorem \cite{Lovelock:1971yv} states, in a four-dimensional spacetime, GR is the most general covariant and local theory of gravity constructed solely from the metric tensor (and associated curvature tensors) that yields second-order field equations. Therefore, any attempt to address the limitations of GR necessarily requires relaxing at least one of the assumptions underlying Lovelock's theorem \cite{Heisenberg_2019}.

Currently, it is unclear which, if any, modified theory of gravity can provide the high-energy corrections required to eliminate singularities, while simultaneously avoiding other pathologies, and identifying such theories is notoriously challenging.
This difficulty in identifying theories where singularities are resolved underlies most of the literature on regular black holes, which focuses purely on studying the properties of black hole geometries that do not present singularities and are geodesically complete \cite{Lan:2023cvz,Carballo-Rubio:2025fnc,Torres:2022twv,1968qtr..conf...87B,Dymnikova:1992ux,Hayward:2005gi,Bambi:2013ufa,Simpson:2019mud,Carballo-Rubio:2023mvr,Carballo-Rubio:2022nuj,Carballo-Rubio:2022kad,DiFilippo:2022qkl,Carballo-Rubio:2021bpr,Carballo-Rubio:2019fnb,Carballo-Rubio:2018pmi,Borissova:2025msp,DiFilippo:2024spj,Simpson:2021dyo,Arrechea:2025nlq,Coviello:2025pla,Franzin:2023slm,Franzin:2022wai}.

However, in recent years there has been great progress in the direction of identifying theories with regular black holes as solutions, mostly within the framework of \emph{Quasi-Topological gravities} \cite{aguilar-gutierrezAspectsHighercurvatureGravities2023, ahmedQuintessentialQuarticQuasitopological2017, buenoFourdimensionalBlackHoles2017, buenoGeneralizedQuasitopologicalGravities2019, buenoGeneralizedQuasitopologicalGravities2022, buenoUniversalBlackHole2017, hennigarBlackHolesEinsteinian2016, hennigarGeneralizedQuasitopologicalGravity2017, morenoClassificationGeneralizedQuasitopological2023a,Moreno:2023arp}\footnote{See also Refs. \cite{Ayon-Beato:1998hmi,Bronnikov:2000vy,Ayon-Beato:2000mjt,Dymnikova:2004zc,Bronnikov:2017sgg,Cano:2020qhy, Cano:2020ezi, Babichev:2020qpr, Baake:2021jzv, Colleaux:2017ibe, buenoRegularBlackHoles2021, buenoRegularChargedBlack2025,Giacchini:2024exc} for other approaches that attempt to construct regular black holes from an action principle.}. This class of theories is constructed such that the field equations in spherically symmetric backgrounds reduce to (at most) second order, and can be integrated into a single algebraic equation. Within this framework, the resolution of the singularity problem in particular settings was achieved in Ref. \cite{Bueno:2024dgm}, where it was shown that an if an infinite tower of higher-curvature corrections to GR is considered, regular black holes naturally arise as the unique spherically symmetric solutions of the theory, and dynamically as the endpoint of gravitational collapse \cite{Bueno:2024zsx,Bueno:2024eig,Bueno:2025gjg}. This approach has since generated a large body of work, see e.g. Refs. \cite{Aguayo:2025xfi,Hennigar:2025ftm,Konoplya:2024hfg,DiFilippo:2024mwm,Konoplya:2024kih,Cisterna:2024ksz,Frolov:2024hhe,Srivastava:2025rxe,Bueno:2025gjg,Konoplya:2025uta,Arbelaez:2025gwj,Bueno:2024zsx,Bueno:2024eig,Bueno:2025gjg,Fernandes:2025fnz,Fernandes:2025eoc}. The main drawback of this approach is, however, that it is only valid in a number of spacetime dimensions $D\geq 5$. We note that this approach was very recently generalized to four dimensions in Ref. \cite{Bueno:2025zaj} by considering non-polynomial curvature terms in the action, i.e. terms schematically of the form $\sim 1/\mathrm{curvature}$. It remains an open question whether these non-polynomial theories are dynamically viable since their actions might become highly singular due to vanishing denominators.

One of the best motivated extensions of GR is \emph{Lovelock gravity}: the unique purely metric-based modified gravity theory that naturally extends GR in higher dimensions by including higher-order curvature terms while maintaining healthy second-order equations of motion \cite{Lovelock:1971yv}. This framework inherently fits with the EFT paradigm and induces higher-order curvature corrections to GR that become particularly relevant in the ultraviolet (UV) regime.
The main issue is that, in four spacetime dimensions, Lovelock's theorem forbids this beyond-GR extension, as all Lovelock terms either become topological or trivial \cite{Fernandes:2022zrq}. Even in higher-dimensions, there are only finitely many Lovelock invariants for a given spacetime dimensionality, which prevents a construction similar to that of quasi-topological gravities.

To bypass Lovelock's theorem, and to construct a Lovelock-corrected four-dimensional theory, several approaches have been proposed, starting with the seminal work of Ref. \cite{Glavan:2019inb}. These attempts started from quadratic Lovelock gravity featuring the Gauss-Bonnet (GB) term, where a non-trivial Einstein-Gauss-Bonnet theory in four dimensions (4DEGB) could be achieved through consistent regularization schemes involving a singular rescaling of the GB coupling. The resulting well-defined theories featured an additional scalar degree of freedom (DOF) beyond the two metric polarisations and hence fall within the \textit{Horndeski class} of scalar-tensor theories, as first uncovered by Refs. \cite{Fernandes:2020nbq, Hennigar:2020lsl, Lu:2020iav, Kobayashi:2020wqy, Fernandes:2021dsb,Bonifacio:2020vbk}, see Ref. \cite{Fernandes:2022zrq} and references within for a comprehensive review of 4DEGB.

Aware of attempts to construct a regularized 4DEGB theory, and of the approach of Ref. \cite{Bueno:2024dgm} to solve the singularity problem, Ref. \cite{Fernandes:2025fnz} considered the regularization scheme of Refs. \cite{Fernandes:2020nbq, Hennigar:2020lsl} applied to the whole tower of Lovelock invariants, leading to particular four-dimensional Horndeski theories. When such an infinite tower of regularized Lovelock-Horndeski corrections is considered, these theories naturally replace the initial Big Bang singularity present in GR by an inflationary period, and also cure the singularity in a particular class of black hole solutions, namely planar black holes, which are also known as ``black branes". Other works have since explored this approach to resolving the singularity problem, see e.g. \cite{Tsujikawa:2025eac,Cisterna:2025vxk,DeFelice:2025fzv,Ling:2025ncw}, and also Ref. \cite{Fernandes:2025eoc} for an extension to $2+1$-dimensional gravity.

The regularization scheme performed to obtain four-dimensional regularized Lovelock theories is not unique; recently, Ref. \cite{Charmousis:2025jpx} considered a different regularization scheme that instead resulted in a four-dimensional theory of 4DEGB belonging to the \textit{generalized Proca class} of vector-tensor theories~\cite{Heisenberg:2014rta}. Interestingly, the static spherically symmetric black holes of the resulting generalized Proca theory studied in Ref. \cite{Charmousis:2025jpx} have primary hair: an independent and free integration constant characterising the black hole in addition to the mass. Due to a number of interesting properties, this Proca 4DEGB theory has been the subject of recent works \cite{Eichhorn:2025pgy,Lutfuoglu:2025ldc,Alkac:2025zzi,Liu:2025dqg,Konoplya:2025uiq,Alkac:2025jhx,Lutfuoglu:2025qkt,Konoplya:2025bte}.
Amongst these works, yet another regularization scheme of the Gauss-Bonnet term was proposed in Ref. \cite{Eichhorn:2025pgy}. This regularization considers two Proca fields, instead of just one, and the resulting four-dimensional bi-Proca theory remarkably allows for regular black holes of the Hayward and gravastar types, providing one of the very few examples where regular black holes are obtained from a four-dimensional action principle. In this case, the primary hair regularizes the singularity, and it can be adjusted such that black holes of all masses can me made extremal, and consequently become immune to the mass-inflation instability that plagues most regular black holes \cite{PhysRevLett.63.1663,PhysRevLett.67.789,Hamilton:2008zz,Brown:2011tv,Carballo-Rubio:2018pmi,Bertipagani:2020awe,Bonanno:2020fgp,Carballo-Rubio:2021bpr,Carballo-Rubio:2022kad,Carballo-Rubio:2024dca}. 

In this paper, in light of the recent work in Proca 4DEGB theories, and in view of infinite towers of corrections as a mechanism to cure singularities, we combine the two approaches. Our main goal is to extend the scalar-tensor tower of regularized Lovelock corrections to the generalized Proca case, and study the resulting phenomenology.

The paper is structured as follows. In Sec. \ref{sec2}, we present the regularized Lovelock framework. Sec. \ref{sec2A} reviews the scalar–tensor regularization of 4DEGB theory and its extension to an infinite Lovelock-Horndeski tower of corrections. Sec. \ref{sec2B} extends the construction to the vector-tensor case, reviews the 4DEGB regularization within Weyl geometry, and introduces the 4D Lovelock-Proca tower action. In Sec. \ref{sec3}, we explore applications of the Lovelock-Proca tower to specific spacetimes: Sec. \ref{sec3A} focuses on cosmology, deriving the generalized Friedmann equations, while Sec. \ref{sec3B} addresses planar black holes. Sec. \ref{sec4} analyzes spherically symmetric black holes. Finally, we conclude in Sec. \ref{sec5}.
We work in units $c=G=1$.

\section{Regularized Lovelock Corrections}\label{sec2}
To set the stage, we start by reviewing the dimensional regularization procedures performed on the Lovelock invariants to obtain well-defined Horndeski theories, and then discuss the generalized Proca case.

\subsection{Reviewing the scalar-tensor case}\label{sec2A}
The $n$-th order Lovelock invariant given by
\begin{equation}
\label{Rn}
    \mathcal{R}^{(n)}\equiv \frac{1}{2^{n}}\delta^{\mu_{1}\nu_{1}...\mu_{n}\nu_{n}}_{\alpha_{1}\beta_{1}...\alpha_{n}\beta_{n}}\prod_{i=1}^{n}R^{\alpha_{i}\beta_{i}}_{\phantom{\alpha_{i}} \phantom{\beta_{i}} \mu_{i}\nu_{i}}, 
\end{equation}
where $\delta^{\mu_{1}\nu_{1}...\mu_{n}\nu_{n}}_{\alpha_{1}\beta_{1}...\alpha_{n}\beta_{n}}\equiv n!\delta^{\mu_{1}\nu_{1}...\mu_{n}\nu_{n}}_{\left [\alpha_{1}\beta_{1}...\alpha_{n}\beta_{n}\right]}$ is the generalized Kronecker delta, and $n\ge 0$. Lovelock gravity extends GR by summing over these invariants in the gravitational action
\begin{equation}
\label{Llovelock}
    \mathcal{L}=\sum_{n=0}^{i}\alpha_{n} \mathcal{R}^{(n)},
\end{equation}
where $\alpha_n$ denotes the coupling constant associated with the $n$-th order curvature invariant. In particular, $n=0$ gives the cosmological constant term, $n=1$ reproduces the Einstein-Hilbert term, and $n=2$ introduces the GB invariant. The index $i$ labelling the highest order of summation from the $D$-dimensional Lagrangian in Eq. \eqref{Llovelock} is given by $i=\frac{1}{2}(D-2)$ for even $D$, and $n=\frac{1}{2}(D-1)$ for odd $D$, since higher-order Lovelock invariants either become topological or vanish. Lovelock theories of gravity coincide with GR plus a cosmological constant in $D=3$ and $D=4$. As a consequence, there are no purely gravitational Lovelock extensions of GR in four-dimensions.

In recent years, progress has been made on extracting non-trivial four-dimensional dynamics from Lovelock densities via controlled dimensional regularization procedures. As an illuminating example, we consider the 4DEGB class of theories, where non-trivial GB-type dynamics are obtained by introducing a divergence in the GB coupling parameter $\alpha_2$ as the spacetime dimension approaches $D\to4$, and by adding appropriate counterterms that prevent divergences in the action as the $D\to 4$ limit is taken. The original approach to 4DEGB by Ref. \cite{Glavan:2019inb} does not consider such counterterms, and consequently, the resulting theory is ill-defined \cite{Gurses:2020ofy}. The divergences arising in the proposal of Ref. \cite{Glavan:2019inb} are related to the conformally invariant terms in the field equations that appear through the Weyl tensor. Therefore, they can be cancelled exactly by adding a conformally rescaled GB term to the action. As we review below, this seemingly ad-hoc regularization scheme does lead to well-defined and physical modified gravity theories: Following closely Refs. \cite{Fernandes:2020nbq,Hennigar:2020lsl}, we hence consider two conformally related metrics
\begin{equation}
    \tilde{g}_{\mu\nu}=e^{2\phi}g_{\mu\nu},
    \label{conformal}
\end{equation}
where $\phi$ is an arbitrary scalar field. The GB invariant of the metric $\tilde{g}_{\mu\nu}$ can be expressed as
\begin{widetext}
\begin{equation}
\label{GBWeyl1}
    \sqrt{-\tilde{g}}\tilde{\mathcal{R}}^{(2)}=e^{(D-4)\phi} \sqrt{-g} \left[\mathcal{R}^{(2)}+(D-3)\nabla_{\mu}J^{\mu}+(D-3)(D-4)\mathfrak{L} \right],
\end{equation}
with
\begin{align}
\label{GBWeyl2}   
&J^{\mu}=8G^{\mu\nu}\nabla_{\nu} \phi+4(D-2)\left[\nabla^{\mu} \phi((\partial \phi)^2+\Box \phi)-\nabla_{\nu} \phi\nabla^{\nu}\nabla^{\mu} \phi\right],\\
\label{GBWeyl3}
&\mathfrak{L}=4G^{\mu\nu}\nabla_{\mu} \phi\nabla_{\nu} \phi+(D-2)\left[4(\partial \phi)^2 \Box \phi+(D-1)(\partial \phi)^4)\right],
\end{align}
\end{widetext}
and we consider the limit
\begin{equation}
    \sqrt{-g}\mathcal{L}^{(2)} = \lim_{D\to 4} \left( \frac{\sqrt{-g} \mathcal{R}^{(2)}}{D-4} - \frac{\sqrt{-\tilde{g}} \tilde{\mathcal{R}}^{(2)}}{D-4} \right),
    \label{eq:regGBST}
\end{equation}
where the first term on the right-hand-side corresponds to the original proposal by Ref. \cite{Glavan:2019inb}, and the second term to a counterterm that removes all divergences associated with the conformally invariant terms and renders the 4D limit well-defined. Using the expressions in Eqs. \eqref{GBWeyl1}-\eqref{GBWeyl3}, we find that the regularized GB term $\mathcal{L}^{(2)}$ is, in fact, part of the Horndeski class of scalar-tensor theories
\begin{equation}
    \begin{aligned}
		&\mathcal{L}_{\mathrm{Horndeski}} = G_2(\phi,X)-G_3(\phi,X)\Box\phi + G_4(\phi,X)R
		\\&
        +G_{4,X}  \left[(\Box\phi)^2-\left(\nabla_\mu \nabla_\nu \phi\right)^2\right]
          +G_5(\phi,X) G^{\mu\nu}\nabla_\mu \nabla_\nu \phi
		  \\&
          -\frac{G_{5,X}}{6}\left[
          (\Box\phi)^3-3\Box\phi \left(\nabla_\mu \nabla_\nu \phi\right)^2
          +2\left(\nabla_\mu \nabla_\nu \phi\right)^3
          \right],
    \end{aligned}
	\label{Horndeski}
\end{equation}
where $X\equiv-\frac{1}{2}(\nabla \phi)^2$, with coupling functions given by
\begin{equation}
    G_2 = 8X^2, \quad G_3 = -8X, \quad G_4 = 4X,\quad G_5 = -4\log X,
\end{equation}
and $G_{i,X}$ denoting a derivative w.r.t. $X$.
The 4DEGB theory is finally obtained by including this regularized GB correction to GR in the gravitational action
\begin{equation}
    S_{\rm 4DEGB} = \frac{1}{16\pi} \int \dd^4x \sqrt{-g}\left( -2\Lambda + R + \alpha_2\mathcal{L}^{(2)} \right),
\end{equation}
The phenomenology of this theory has been widely studied and we refer the reader to Ref. \cite{Fernandes:2022zrq} for a comprehensive review.

The regularization method applied to the GB term to obtain a well-defined scalar–tensor GB theory can be generalized systematically to all orders of Lovelock invariants by taking the limit
\begin{equation}
\label{limLn}
    \mathcal{L}^{(n)}=\lim_{D\to2n}\frac{\sqrt{-g}\mathcal{R}^{(n)}-\sqrt{-\tilde{g}}\tilde{\mathcal{R}}^{(n)}}{d-2n},
\end{equation}
where we note that the $n$-th order Lovelock invariant $\mathcal{R}^{(n)}$ becomes topological as the spacetime dimension $D$ approaches the critical dimension $2n$. After this limit is taken, the resulting Lagrangian should be evaluated inside a four-dimensional action to obtain a four-dimensional theory. The Lagrangians in Eq. \eqref{limLn} were studied in Refs. \cite{Colleaux:2020wfv,Fernandes:2025fnz} and give rise to Horndeski theories at each order $n$, with coupling functions given by
\begin{equation}
\label{HorndeskiCoupling4d}
\begin{aligned}
    &G_{2}^{(n)}=2^{n+1}(n-1)(2n-3)X^{n},\\
    &G_{3}^{(n)}=-2^{n}n(2n-3)X^{n-1},\\
    &G_{4}^{(n)}=2^{n-1}nX^{n-1},\\
    &G^{(n)}_5 = -\begin{cases} 
      4 \log X, & n = 2, \\ 
      2^{n-1} \frac{n(n-1)}{n-2} X^{n-2}, & n > 2,
   \end{cases}
\end{aligned}
\end{equation}
which leads to the following Lovelock-corrected theory of gravity in four dimensions
\begin{equation}
\label{Sst4d}
    S=\frac{1}{16\pi}\int \dd^{4}x\sqrt{-g}\left[-2\Lambda+R+\frac{1}{\ell^{2}}\sum_{n=2}^{\infty}c_{n}\ell^{2n}\mathcal{L}^{(n)}_{\mathrm{ST}}\right],
\end{equation}
where $\{c_{n}\}$ are dimensionless coupling constants and $\ell$ denotes the  length scale controlling the beyond-GR effects. \\

In general, Horndeski theories exhibit a global shift symmetry $\phi\to\phi+C$ for any constant $C$, resulting in a conserved Noether current $j^{\mu}$, with $\nabla_{\mu}j^{\mu}=0$ being equivalent to the scalar field equation of motion. Evidently, imposing $j^{\mu}=0$ constitutes a possible solution, for which various cosmological and static planar black hole solutions have been obtained. In these scenarios, the Big Bang singularity is naturally replaced by an inflationary de Sitter phase, with a smooth exit to standard GR at late times, while static planar black hole solutions become regular. In these solutions, regularity is lost as soon as one truncates the curvature corrections at any specific order. One of the shortcomings of this approach is that no (analytic) spherically symmetric black hole solutions have yet been found.

In this work, similar solutions will be explored for the class of generalized Proca theories, where the generalized Proca Lagrangians are obtained from an infinite tower of dimensionally regularized Lovelock corrections based on Weyl geometry. The construction of the vector-tensor theories of interest will be discussed in the next section.

\subsection{Generalized Proca case}\label{sec2B}
Similar to the scalar-tensor theories discussed above, regularized Lovelock corrections can also give rise to vector-tensor theories. The vector field originates from generalizing the regularization procedure used in the scalar-tensor case by considering the framework of Weyl geometry, which naturally relies on a vector field $A_{\mu}$ to break the metricity condition
\begin{equation}
\label{WeylGeometry}
    \hat{\nabla}_{\lambda}g_{\mu\nu}=-2g_{\mu\nu}A_{\lambda},
\end{equation}
where the covariant derivative $\hat{\nabla}_{\mu}$ is associated to the Weyl connection
\begin{equation}
\label{WeylConnection}   \hat{\Gamma}^{\lambda}_{\,\,\,\mu\nu}=\Gamma^{\lambda}_{\,\,\,\mu\nu}+\delta^{\lambda}_{\,\,\,\mu}A_{\nu}+\delta^{\lambda}_{\,\,\,\nu}A_{\mu}-g_{\mu\nu}A^{\lambda},
\end{equation}
with $\Gamma^{\lambda}_{\,\,\,\mu\nu}$ denoting the metric-compatible Levi-Civita connection (see also \cite{BeltranJimenez:2016wxw}).

The dimensional regularization of Lovelock invariants can be performed analogously to Eq. \eqref{limLn}, where Lovelock invariants constructed from the Weyl connection are added as counterterms. In particular, this regularization procedure has recently been performed on the Gauss-Bonnet term in Ref. \cite{Charmousis:2025jpx}, where the GB invariant constructed from the Weyl connection can be written as \cite{BeltranJimenez:2014iie,Bahamonde:2025qtc}
\begin{equation}
\label{GBWeyl1P}
    \hat{\mathcal{G}}=\mathcal{G}+(d-3)\nabla_{\mu}J^{\mu}+(d-3)(d-4)\mathfrak{L},
\end{equation}
with
\begin{align}
\label{GBWeyl2P}   
&J^{\mu}=8G^{\mu\nu}A_{\nu}+4(d-2)\left[A^{\mu}(A^{2}+\nabla_{\nu}A^{\nu})-A_{\nu}\nabla^{\nu}A^{\mu}\right],\\
&\mathfrak{L}=4G^{\mu\nu}A_{\mu}A_{\nu}+(d-2)\left[4A^{2}\nabla_{\mu}A^{\mu}+(d-1)A^{4}\right],
\end{align}
where we have abbreviated $A^{2}=A_{\mu}A^{\mu}$ and $A^{4}=(A^{2})^{2}$.

Analogous to Eq. \eqref{eq:regGBST}, the regularized four-dimensional Proca-GB theory of Ref. \cite{Charmousis:2025jpx} can be obtained through the following regularization 
\begin{equation}
\label{DimRegVT4DGB}
\mathcal{L}_{\mathcal{G}}^{\text{VT}}=\lim_{D\to4}\frac{\mathcal{G}-\hat{\mathcal{G}}}{D-4},
\end{equation}
which gives rise to the following vector-tensor theory
\begin{equation}
\label{L4DGBVT}    \mathcal{L}_{\mathcal{G}}^{\text{VT}}=-4G^{\mu\nu}A_{\mu}A_{\nu} - 8A^{2}\nabla_{\mu}A^{\mu} - 6A^{4},
\end{equation}
up to total derivatives, leading to a 4DEGB Proca theory given by
\begin{equation}
\label{S4DEGBVT}
    S=\int d^{4}x\sqrt{-g}(R+\alpha_2\mathcal{L}_{\mathcal{G}}^{\text{VT}}).
\end{equation}
Similar to scalar-tensor 4DEGB theory, the resulting theory is a specific case of the class of generalized Proca theories \cite{Heisenberg:2014rta, Heisenberg_2019}
\begin{equation}
\label{Lgp}
    \mathcal{L}_{\text{gen.Proca}}=\sum_{n=2}^{6}\mathcal{L}_{n},
\end{equation}
with the following Lagrangians
\begin{equation}\label{eq:Lall}
\begin{aligned}
\mathcal{L}_2 &= G_2\!\left(A_\mu,\,F_{\mu\nu},\,\tilde F_{\mu\nu}\right),\\[4pt]
\mathcal{L}_3 &= G_3(X)\,\nabla_\mu A^\mu,\\[4pt]
\mathcal{L}_4 &= G_4(X)\,R
+ G_{4,X}\!\left[\,(\nabla_\mu A^\mu)^2-\nabla_\rho A_\sigma\,\nabla^\sigma A^\rho\right],\\[4pt]
\mathcal{L}_5 &= G_5(X)\,G_{\mu\nu}\,\nabla^\mu A^\nu
-\frac{1}{6}G_{5,X}\Big[(\nabla\!\cdot\!A)^3 \\ 
&\quad + 2\,\nabla_\rho A_\sigma\,\nabla^\gamma A^\rho\,\nabla_\gamma A^\sigma
- 3\,(\nabla\!\cdot\!A)\,\nabla_\rho A_\sigma\,\nabla^\sigma A^\rho\Big]\\ 
&\quad - g_5(X)\,\tilde F^{\alpha\mu}\,\tilde F^{\beta}{}_{\mu}\,\nabla_\alpha A_\beta,\\[4pt]
\mathcal{L}_6 &= G_6(X)\,\mathcal{L}^{\mu\nu\alpha\beta}\,\nabla_\mu A_\nu\,\nabla_\alpha A_\beta\\
&\quad+\frac{G_{6,X}}{2}\,\tilde F^{\alpha\beta}\tilde F^{\mu\nu}\,\nabla_\alpha A_\mu\,\nabla_\beta A_\nu,
\end{aligned}
\end{equation}
where $X\equiv-A_{\mu}A^{\mu}/2$ denotes the vector field norm, $F_{\mu\nu}\equiv\nabla_{\mu}A_{\nu}-\nabla_{\nu}A_{\mu}$ the field strength tensor with corresponding dual $\tilde{F}_{\mu\nu}$, and $\mathcal{L}^{\mu\nu\alpha\beta}$ the double-dual Riemann tensor. The 4DEGB Proca Lagrangian in Eq. \eqref{L4DGBVT} corresponds to a specific generalized Proca theory with the following coupling functions
\begin{equation}
\label{couplingGP4DEGB}
    G_{2}=-24X^{2},\quad G_{3}=-16X, \quad G_{4}=-4X,
\end{equation}
and $G_5,\,G_6$ are vanishing. While there is no term analogous to $G_6$ in the Horndeski Lagrangian, the vanishing of the $G_5$-function is a crucial difference between the scalar- and vector-tensor theories, as even in the limit where the dynamics are dominated by the longitudinal mode $A_\mu \rightarrow \nabla_\mu \phi$, the resulting Horndeski theories are fundamentally distinct.
From a technical point of view, the $G_5$-term is absent in the vector-tensor case because $\sqrt{-g}$ does not play a role in the regularization, while in the scalar-tensor case, the theory picks up a term $\sim \phi \mathcal{R}^{(2)} \subset G_5$ when expanding the exponential in Eq. \eqref{GBWeyl1} around $D=4$.
No kinetic term $\sim F_{\mu \nu} F^{\mu \nu}$ shows up in the dimensional regularization procedure, which we comment on later.

Following the spirits of regularized Lovelock-Horndeski theories and 4DEGB Proca theory, we now consider GR corrected by an infinite tower of higher-order Lovelock-Proca terms
 \begin{equation}
    S = \frac{1}{16\pi} \int d^4x \sqrt{-g} \left[ -2\Lambda + R + \frac{1}{\ell^2} \sum_{n=2}^{\infty} c_n \ell^{2n} \mathcal{L}^{(n)}_{\text{VT}}\right],
\label{S4d}
\end{equation}
where $\mathcal{L}^{(n)}_{\text{VT}}$ denotes the generalized Proca Lagrangians in Eq. \eqref{eq:Lall} equipped with the following coupling functions
\begin{align}
\label{proca4d}
    &G_2^{(n)}(X)=-6\cdot2^{n}(n-1)X^n,\notag\\
    &G_3^{(n)}(X)=-2^{n+1}nX^{n-1},\\
    &G_4^{(n)}(X)=-2^{n-1}\frac{n}{2n-3}X^{n-1},\notag
\end{align}
and $G_5^{(n)}=G_6^{(n)}=0$.
When $n=2$, we evidently recover the 4DEGB Proca case \eqref{couplingGP4DEGB}, while higher-order corrections are obtained for higher $n$. The derivation of Eq. \eqref{proca4d} follows from considering the expressions obtained in Ref. \cite{Colleaux:2020wfv} for the scalar-tensor case, and noticing that in that case, upon a conformal transformation, the connection changes as in Eq. \eqref{WeylConnection} upon setting $A_\mu \rightarrow \nabla_\mu \phi$. Then, all expressions from Ref. \cite{Colleaux:2020wfv} can be used in the Proca case, upon identifying $\nabla_\mu \phi = A_\mu$, and ignoring the overall exponential factor that does not appear in the Proca case since the metric is not transformed, only the connection. The absence of this exponential factor has important consequences, as it directly implies that the resulting generalized Proca theory does \emph{not} have a $G_5$-term.

When the dynamics of the theory are dominated by the longitudinal mode $A_\mu \rightarrow \nabla_\mu \phi$, the theory reduces to a shift-symmetric Horndeski theory described by the same coupling functions. Nonetheless, and importantly, much like in the GB case, even in this limit, the infinite tower of Lovelock-Proca corrections \eqref{proca4d} and the infinite tower of Lovelock-Horndeski corrections \eqref{HorndeskiCoupling4d} describe physically distinct theories, as their associated coupling functions are different, and crucially, there is no $G_5$-term present in the Proca case. On a physical level, this difference is made explicit in Sec. \ref{sec4}, where we find spherically symmetric black hole solutions exhibiting primary Proca-hair from the Proca-Lovelock tower \eqref{S4d}.

\section{Singularity resolution from an infinite tower of Proca corrections}\label{sec3}
In this section, we are interested in analysing the phenomenological consequences of the regularized Proca-Lovelock tower in particular spacetimes; first, investigating the resulting dynamics on a cosmological Friedmann-Lemaître-Robertson-Walker (FLRW) background, and secondly, exploring black hole solutions.

\subsection{Cosmology and inflation}\label{sec3A}
We consider a spatially flat FLRW metric 
\begin{equation}
\label{frw}
    \dd s^2=-\dd t^2+a(t)^2(\dd r^2+r^2\dd \theta^2+r^2\sin^2\theta\,\dd \phi^2),
\end{equation}
where $a(t)$ denotes the scale factor, and the Hubble parameter is defined as $H=\frac{\dot{a}}{a}$. We consider a perfect fluid stress-energy tensor for matter $T^{\mu}_{\,\,\nu}=\mathrm{diag}(-\rho,p,p,p)$, where the matter energy density $\rho$ and pressure $p$ satisfy the continuity equation $\dot{\rho}+3H(\rho+p)=0$.\\

We further assume the vector field ansatz
\begin{equation}
    A_{\mu}dx^{\mu}=A_{0}(t)dt,
\end{equation}
compatible with the homogeneity and isotropy of the background. The equations of motion for generalized Proca theories on the geometry \eqref{frw} were already computed in \cite{Heisenberg_2019} (see also \cite{DeFelice:2016yws, DeFelice:2016uil, deFelice:2017paw, Heisenberg:2020xak}). At each order $n$, they are given by
 \begin{equation}
 \label{cosmo1}
     \begin{aligned}
         &G^{(n)}_2 - G^{(n)}_{2,X}A_0^2 - 3G^{(n)}_{3,X} H A_0^3 \\&- 6 (2G^{(n)}_{4,X} + G^{(n)}_{4,XX} A_0^2) H^2 A_0^2 + 6 G^{(n)}_4 H^2  = 16\pi \rho,
     \end{aligned}
 \end{equation}
 \begin{equation}
 \label{cosmo2}
     \begin{aligned}
    &G^{(n)}_2 - \dot{A_0}A_{0}^{2} G^{(n)}_{3,X} + 2G^{(n)}_4 \left(3H^2 + 2\dot{H}\right) \\
    &- 2G^{(n)}_{4,X}A_0 \left(3H^2 A_{0}+ 2H\dot{A_0} + 2\dot{H}A_0\right)\\&-4G^{(n)}_{4,XX}H\dot{A_0}A_{0}^{3}=-16\pi p,
\end{aligned}
 \end{equation}
for the background metric and
\begin{equation}
\label{cosmo3} A_0\left(G^{(n)}_{2,X}+3G^{(n)}_{3,X}HA_0+6G^{(n)}_{4,X}H^2+6G^{(n)}_{4,XX}H^2A_0^{2}\right)=0,
\end{equation}
for the vector field, where we have ignored the contributions from the $G_{5}$ terms due to the nature of Lovelock-Proca theories we consider.

For the specific theory Eq. \eqref{proca4d} at hand, the vector-field equation is solved by the following vector profile
\begin{equation}
    A_0(t)= -H(t),
\end{equation}
for all values of $n$.
Using this vector field profile in the Einstein equations, we find the $n$-th order contribution to the first Friedmann equation to be proportional to $H^{2n}$, leading to the generalized Friedmann equation
\begin{equation}
\label{GeneralisedFriedmann}
    F(H^2)\equiv\frac{1}{\ell^2}\sum_{n=1}^{\infty}c_{n}(\ell^2H^2)^n=\frac{8\pi}{3}\rho+\frac{\Lambda}{3},
\end{equation}
where the function $F$ is defined as
\begin{equation}
\label{F}
    F(x)\equiv\frac{1}{\ell^2}\sum_{n=1}^{\infty}c_{n}(\ell^2x)^n
\end{equation}
The generalized Friedmann equation \eqref{GeneralisedFriedmann} lies at the core of resolving the initial singularity: when $\rho\rightarrow \infty$, characterizing the very early universe, the function $F$ in Eq. \eqref{F} gives us the possibility to respect Eq. \eqref{GeneralisedFriedmann} without reaching a curvature singularity ($H^2\to \infty$) as in GR, but by requiring that the function $F(H^2)$ exhibits a pole at finite $H^2$ instead. Sufficient conditions on the coupling functions $c_n$ for the existence of such a pole are given by
\begin{align}
    &(i)\,\, \textrm{Nonnegative coefficients: }\,c_n\geq 0,\quad \forall n\in\mathbb{N},\\[5pt]
    &(ii)\,\,\textrm{Finite and non-zero radius of convergence:}\\[5pt]
    &\hspace{30pt}R\equiv\frac{1}{\textrm{lim}_{n\rightarrow\infty}(c_n)^{1/n}}\hspace{10pt}\textrm{s.t.}\hspace{5pt}0<R<\infty,\\[5pt]
    &(iii)\,\,\textrm{The series diverges at } R:\hspace{5pt}\sum_{n=0}^\infty c_nR^n\rightarrow\infty
\end{align}
Assuming the above conditions hold, the radius of convergence can be chosen $R=1$ without loss of generality (by rescaling all $c_n$ by a constant non-zero value). Then, a pole would exist as $H^2 \to 1/\ell^2$.

There are infinitely many possibilities to choose suitable couplings $\{c_n\}$ that lead to a pole in $F$ in the early universe. For example, Ref. \cite{Fernandes:2025fnz} considered the illustrative example of $c_n=(1-(-1)^n)/(2n)$ to demonstrate explicitly how the initial singularity is cured from the pole structure: This choice of coefficients allows to resum the generalized Friedmann equation to $F(H^2)=\textrm{tanh}^{-1}(\ell^2H^2)/\ell^2$, which incurs a pole as the Hubble parameter approaches a constant value $H^2\rightarrow 1/\ell^2$. Evidently, the Friedmann equation can be recast as follows
\begin{equation}
    H^2=\frac{1}{\ell^2}\textrm{tanh}\,\left[\ell^2\left(\frac{8\pi}{3}\rho+\frac{\Lambda}{3}\right)\right].
\end{equation}
As the energy density diverges, the early universe asymptotes towards a regular de Sitter geometry with $H^2\rightarrow 1/\ell^2$. Given, however, that an infinite energy density cannot be reached in a finite amount of time, the universe starts evolving from a \textit{quasi}-de Sitter expansion directly. Overall, the initial Big Bang singularity is not only resolved by the systematic inclusion of all proposed gravitational UV physics encoded in the Proca-Lovelock tower, but also directly replaced by a period of inflation. Notice that this model is clearly distinct from non-singular bouncing cosmologies, as it does not rely on a contracting phase preceding the inflationary expansion (independently of the stability issues \cite{Bohnenblust:2024mou}). Furthermore, Ref. \cite{Fernandes:2025fnz} has shown explicitly that the inflationary phase is terminated by a smooth transition into a radiation-dominated universe, providing a natural graceful exit to standard GR.

\subsection{Regular planar black holes}\label{sec3B}
Black holes in GR generically contain singularities. In this section, we explore how the modified gravity theory given by Eqs. \eqref{S4d}-\eqref{proca4d} allows for regular black hole solutions. We start by considering a line element of the form
\begin{equation}
\label{lineel}
    \dd s^2=-f(t,r)N(t,r)^2\dd t^2+\frac{\dd r^2}{f(t,r)}+r^2\dd \Omega_{k}^2,
\end{equation}
with $N(t,r)$ the lapse function and $k\in\left \{ -1,0,1 \right \} $ corresponding to positive, zero and negative horizon curvatures. As the resulting field equations take the simplest form for planar geometries, we first focus on the case $k=0$. In GR, planar black hole solutions require a negative cosmological constant $\Lambda=-3/L^2$ expressed in terms of the Anti-de-Sitter (AdS) radius $L$, and they reduce to the following static solution \cite{Birmingham_1999}
\begin{align}
\label{4dfGR}
    &N(t,r)=1,\\
    &f(t,r) \equiv f_{\rm GR}(r)=\frac{r^2}{L^2}-\frac{2M}{r},
\end{align}
with $M$ denoting an integration constant related to the black hole mass. Evidently, the resulting spacetime geometry features a curvature singularity at $r=0$.\\

For the Lovelock-Proca theory defined in Eqs. \eqref{S4d}-\eqref{proca4d}, we assume the following static vector field ansatz
\begin{equation}
\label{VectorAnsatz}
    A_{\mu}dx^{\mu}=A_{0}(r)dt+A_{1}(r)dr.
\end{equation}
Using this vector profile together with the line element \eqref{lineel} in the theory Eq. \eqref{S4d}, we find that the vector field equations of motion factorize and are solved by
\begin{equation}
\label{AprofileBH}
    A_0(r)=0, \quad A_{1}(r)=-\frac{1}{r},
\end{equation}
independently of the order $n$. Notice that this vector profile is purely governed by the longitudinal scalar mode $\phi(r)=\log(r/r_0)$ as found in \cite{Fernandes:2025fnz}, with $r_0$ an arbitrary constant. However, given that the coupling functions in the Proca-Lovelock theory \eqref{proca4d} are different from those in the Horndeski tower \eqref{HorndeskiCoupling4d}, the resulting black hole solutions could still differ from each other. Therefore, we proceed by substituting the vector field profile into the the Einstein equations and directly find that the $(tr)$-component imposes $f(t,r)=f(r)$. Moreover, a suitable combination of the $(tt)$- and $(rr)$-equations constrains $N(t,r)=N(t)$, which amounts to a simple rescaling of the time coordinate $t$. Consequently, we can set $N(t)=1$ without loss of generality.

At this stage, the system reduces to a single differential equation for $f(r)$ at each order $n$ given by
\begin{equation}
    \sum_{n=0}^{\infty} c_n \, \ell^{2n-2} \left( -\frac{f(r)}{r^2} \right)^n \left[ (2n-3)f(r) - nr f'(r) \right] = 0,
    \label{ODE4dBH}
\end{equation}
which can be integrated into an algebraic equation after multiplying by an overfall factor $(\frac{-f(r)}{r^2})^{-1}$, yielding
\begin{equation}
    F\left(-\frac{f(r)}{r}\right) = -\frac{f_{\text{GR}}(r)}{r^2},
    \label{4dBHsol}
\end{equation}
with $f_{\text{GR}}$ as in Eq. \eqref{4dfGR} and the function $F$ defined in \eqref{F} with $c_{1}=1$. Notice that we directly recover the GR solution upon truncating the series at $n=1$.

The sufficient conditions to resolve the planar black hole singularity are the same ones as presented for the cosmological case, and have been discussed in detail in Ref. \cite{Fernandes:2025fnz}, along with particular examples.

\section{Spherically symmetric black holes}\label{sec4}
For spherical horizons with $k=1$ in Eq. \eqref{lineel}, the simplifications that led to integrability in the planar case fail in an essential way. Starting from the ansatz
\begin{align}
\label{4dspheric}
    &\dd s^2=-f(r)N(r)^2 \dd t^2+\frac{\dd r^2}{f(r)}+r^2(\dd \theta^2+\sin ^2\theta^2 \dd\phi^2),\notag\\
    &A_{\mu}\dd x^{\mu}=A_{0}(r)\dd t+A_{1}(r)\dd r,
\end{align}
the universal vector profile \eqref{AprofileBH} that factorizes the Proca equations for $k=0$ no longer solves field equations. This is observed by considering the effective one-dimensional Lagrangian and varying with respect to $N$, $f$, $A_{0}$ and $A_{1}$, which yields a system of equations that does not admit a closed-form non-trivial Proca profile at generic order $n$ in the tower without either forcing a constant norm $X$ or a vanishing vector field. Consequently, without additional assumptions, we were not able to integrate the field equations.

However, when the series is truncated at $n=2$, corresponding to the Proca-GB theory studied in Ref. \cite{Charmousis:2025jpx}, the field equations impose $X = \mathrm{cte}$, and the geometry is independent of the constant value of $X$. In that case, the solution reads \cite{Charmousis:2025jpx}
\begin{equation}
    A_1 = \frac{A_0}{f}, 
    \label{eq:X=0}
\end{equation}
\begin{equation}
    A_0 = \frac{r}{4\ell^2c_2} \left( 1-\sqrt{1+\frac{8\ell^2 Qc_2}{r^3}} \right),
\end{equation}
\begin{equation}
    f = 1-\frac{2(M-Q)}{r} + \frac{r^2}{2\ell^2c_2}\left( 1-\sqrt{1+\frac{8\ell^2 Qc_2}{r^3}} \right),
    \label{eq:PrimaryHairSol1}
\end{equation}
together with $N=1$ (which we assume implicitly from now on), and where we have imposed $X=0$ for simplicity. The solution depends on two integration constants: the ADM mass $M$ of the black hole, and a free primary-hair constant $Q$ (see also \cite{Heisenberg:2017hwb, Heisenberg:2017xda, DeFelice:2016cri} for comparison purposes).

To integrate the field equations in our case for the whole Lovelock-Proca tower, we follow this example and start by assuming $X=0$ for simplicity, which is imposed by the relation in Eq. \eqref{eq:X=0}, while discussing the $X=\mathrm{cte}\neq 0$ later. Under this assumption, and following the method of Ref. \cite{Charmousis:2025jpx}, the field equations for the theory \eqref{S4d} with generic couplings $\{c_n\}$ allow for an exact solution that reads
\begin{equation}
    A_0 = \frac{r c_2}{2\ell^2 (2c_2^2-c_3)}\left( 1-\sqrt{1+\frac{4\ell^2 Q(2c_2^2-c_3)}{r^3c_2}} \right),
\end{equation}
\begin{equation}
\begin{aligned}
    f =& 1-\frac{2\left(M-\frac{2Q c_2^2}{2c_2^2-c_3}\right)}{r} \\&+ \frac{2r^2c_2^3}{\ell^2(2c_2^2-c_3)}\left( 1-\sqrt{1+\frac{4\ell^2 Q(2c_2^2-c_3)}{r^3c_2}} \right).
    \end{aligned}
    \label{eq:PrimaryHairSol2}
\end{equation}
Note that the only terms in the tower of corrections that contribute to the solution are $n=2$ and $n=3$, because for higher orders $n>3$, the field equations are all proportional to a positive power of $X$, which vanishes in this case. The solution of Ref. \cite{Charmousis:2025jpx} is recovered when $c_3=0$. Although the solutions given in \eqref{eq:PrimaryHairSol1} and \eqref{eq:PrimaryHairSol2} are similar in shape, they define distinct geometries, and are solutions to different theories. Therefore, the solution in Eq. \eqref{eq:PrimaryHairSol2} is a genuinely new solution.

In Eq. \eqref{eq:PrimaryHairSol2}, there is a case in the space of couplings that stands out, namely $2c_2^2=c_3$. In this case, the geometry and the profile for the vector field simplify drastically and we get
\begin{equation}
    A_0 = -\frac{Q}{r^2},
\end{equation}
\begin{equation}
    f = 1-\frac{2M}{r} + \frac{4\ell^2 Q^2 c_2}{r^4}.
\end{equation}
Therefore, when $2c_2^2=c_3$, the solution can be expressed solely in powers of $1/r$, much like the Schwarzschild and Reissner-Nordstr\"om metrics, and the vector component $A_0$ falls-off like $\sim 1/r^2$. A geometry similar to this one was previously derived in the context of quantum gravity \cite{Fazzini:2023scu,Rovelli:2024sjl}.

For simplicity, we have so far assumed $X=0$. Nonetheless, the field equations can also be integrated when $X=\mathrm{cte}\equiv \kappa$, which is imposed by taking
\begin{equation}
    A_1 = \frac{\sqrt{A_0^2+2\kappa f}}{f}.
\end{equation}
Because $X$ is non-vanishing, the Lagrangians corresponding to $n>3$ will also contribute. However, in this case, the only solutions we were able to find are not asymptotically flat. As a tractable, simple example, we impose $\{c_n\}=0$, for $n>3$, and restrict to the following couplings and vector norm
\begin{equation}
    c_3 = 2(2-\sqrt{6})c_2^2, \qquad X\equiv \kappa = \frac{3+\sqrt{6}}{12\ell^2c_2},
    \label{eq:couplings-dS-BH}
\end{equation}
as the field equations simplify considerably (similarly to the case $2c_2=c_3^2$ presented above). In this case, a solution to the field equations is given by
\begin{equation}
    A_0 = -\frac{Q}{r^2} + \frac{\left(\sqrt{6}-4 \right)r}{12\ell^2 c_2}
\end{equation}
\begin{equation}
    f = 1-\frac{2M}{r} - \frac{4\sqrt{6}\ell^2 Q^2 c_2}{r^4} - \frac{2561 \sqrt{6}+6276}{1800}\frac{r^2}{\ell^2 c_2},
\end{equation}
which is asymptotically anti-de Sitter or de Sitter, depending on the sign of $c_2$. While we fixed the couplings as in \eqref{eq:couplings-dS-BH} to have expressions which are tractable and presentable, in general, the solution to the field equations is such that the effective cosmological constant is proportional to $\kappa$. Given that $\kappa$ is a free parameter that is not fixed in term of any couplings of the theory, the cosmological constant emerges as an integration constant in this case, which might alleviate the cosmological constant problem.

Although these solutions are not regular, as they contain singularities within the event horizon, they provide yet another distinction with respect to the scalar-tensor case of Ref. \cite{Fernandes:2025fnz}, where spherically-symmetric exact solutions could not be found when considering Horndeski couplings at orders $n>2$. We refer to \cite{Heisenberg:2017hwb, Heisenberg:2017xda, DeFelice:2016cri} for similar interesting hairy solutions of generalized Proca theories.

\section{Conclusions}\label{sec5}
In this work, we have addressed the problem of spacetime singularities, an inherent prediction of GR appearing at the centre of black holes and the origin of the universe, by including an infinite tower of higher-order Proca corrections in the gravitational action. Since Lovelock's theorem prevents any Lovelock theory in four dimensions, we applied a regularization scheme based on Weyl geometry to each curvature invariant in order to obtain non-vanishing corrections. This regularization scheme was first introduced in Ref. \cite{Charmousis:2025jpx} and applied to the Gauss-Bonnet term, resulting in a 4DEGB theory belonging to the generalized Proca class of vector-tensor theories. 
Our work extended this analysis to higher-order Lovelock invariants, leading to a whole tower of generalized Proca theories in four dimensions. This Lovelock-Proca tower replaces the initial Big Bang singularity by a regular de Sitter phase, giving us an inflationary period at the beginning of the universe for free, with a graceful exit to standard GR evolution. Furthermore, the Lovelock-Proca corrections allow for the construction of regular planar black hole solutions in four dimensions. 
In contrast to the scalar-tensor case discussed in \cite{Fernandes:2025fnz}, we also find spherically symmetric black hole solutions endowed with primary hair, which emerges as a free integration constant that is independent of the theory's couplings constants and of other global charges such as the ADM mass. One set of hairy black hole solutions follows from having a vanishing vector field norm, which allows only the Lovelock invariants up to $n=3$ to contribute and hence the resulting geometries are not regular. The second set of solutions stems from having a constant vector field norm, which a priori includes contributions from Lagrangians $\mathcal{L}_{n>3}$, but leads to geometries that are not asymptotically flat. Our findings underline the importance of including the full tower of corrections as a means to cure spacetime singularities, a feature that is shared by many leading candidates for quantum gravity, such as string theory and asymptotically safe quantum gravity.  

Although the standard kinetic term for the Proca field $\sim F_{\mu \nu} F^{\mu \nu}$ was not included, it can be added without affecting the analysis in Sec. \ref{sec3}, as only the longitudinal mode contributes in these cases. This contrasts with the scalar–tensor scenario, where a canonical kinetic term disrupts integrability. While the kinetic term does not naturally emerge from the dimensional regularization procedure, it may be necessary to ensure well-behaved perturbations and to avoid strong coupling \cite{Tsujikawa:2025eac,DeFelice:2025fzv}.

Regarding future directions, it would be interesting to further analyse the newly found hairy black hole solutions, as primary hair can have drastic effects on the resulting geometries. One clear path ahead lies in studying the quasi-normal modes of the black holes that are characteristically altered by the presence of the primary hair and therefore serve as an important observational avenue in testing modified theories of gravity. 

\bibliography{references}

\begin{thebibliography}{122}%
\makeatletter
\providecommand \@ifxundefined [1]{%
 \@ifx{#1\undefined}
}%
\providecommand \@ifnum [1]{%
 \ifnum #1\expandafter \@firstoftwo
 \else \expandafter \@secondoftwo
 \fi
}%
\providecommand \@ifx [1]{%
 \ifx #1\expandafter \@firstoftwo
 \else \expandafter \@secondoftwo
 \fi
}%
\providecommand \natexlab [1]{#1}%
\providecommand \enquote  [1]{``#1''}%
\providecommand \bibnamefont  [1]{#1}%
\providecommand \bibfnamefont [1]{#1}%
\providecommand \citenamefont [1]{#1}%
\providecommand \href@noop [0]{\@secondoftwo}%
\providecommand \href [0]{\begingroup \@sanitize@url \@href}%
\providecommand \@href[1]{\@@startlink{#1}\@@href}%
\providecommand \@@href[1]{\endgroup#1\@@endlink}%
\providecommand \@sanitize@url [0]{\catcode `\\12\catcode `\$12\catcode `\&12\catcode `\#12\catcode `\^12\catcode `\_12\catcode `\%12\relax}%
\providecommand \@@startlink[1]{}%
\providecommand \@@endlink[0]{}%
\providecommand \url  [0]{\begingroup\@sanitize@url \@url }%
\providecommand \@url [1]{\endgroup\@href {#1}{\urlprefix }}%
\providecommand \urlprefix  [0]{URL }%
\providecommand \Eprint [0]{\href }%
\providecommand \doibase [0]{https://doi.org/}%
\providecommand \selectlanguage [0]{\@gobble}%
\providecommand \bibinfo  [0]{\@secondoftwo}%
\providecommand \bibfield  [0]{\@secondoftwo}%
\providecommand \translation [1]{[#1]}%
\providecommand \BibitemOpen [0]{}%
\providecommand \bibitemStop [0]{}%
\providecommand \bibitemNoStop [0]{.\EOS\space}%
\providecommand \EOS [0]{\spacefactor3000\relax}%
\providecommand \BibitemShut  [1]{\csname bibitem#1\endcsname}%
\let\auto@bib@innerbib\@empty
\bibitem [{\citenamefont {Adame}\ \emph {et~al.}(2025)\citenamefont {Adame} \emph {et~al.}}]{DESI:2024mwx}%
  \BibitemOpen
  \bibfield  {author} {\bibinfo {author} {\bibfnamefont {A.~G.}\ \bibnamefont {Adame}} \emph {et~al.} (\bibinfo {collaboration} {DESI}),\ }\bibfield  {title} {\bibinfo {title} {{DESI 2024 VI: cosmological constraints from the measurements of baryon acoustic oscillations}},\ }\href {https://doi.org/10.1088/1475-7516/2025/02/021} {\bibfield  {journal} {\bibinfo  {journal} {JCAP}\ }\textbf {\bibinfo {volume} {02}},\ \bibinfo {pages} {021}},\ \Eprint {https://arxiv.org/abs/2404.03002} {arXiv:2404.03002 [astro-ph.CO]} \BibitemShut {NoStop}%
\bibitem [{\citenamefont {Abbott}\ \emph {et~al.}(2024)\citenamefont {Abbott} \emph {et~al.}}]{DES:2024jxu}%
  \BibitemOpen
  \bibfield  {author} {\bibinfo {author} {\bibfnamefont {T.~M.~C.}\ \bibnamefont {Abbott}} \emph {et~al.} (\bibinfo {collaboration} {DES}),\ }\bibfield  {title} {\bibinfo {title} {{The Dark Energy Survey: Cosmology Results with {\ensuremath{\sim}}1500 New High-redshift Type Ia Supernovae Using the Full 5 yr Data Set}},\ }\href {https://doi.org/10.3847/2041-8213/ad6f9f} {\bibfield  {journal} {\bibinfo  {journal} {Astrophys. J. Lett.}\ }\textbf {\bibinfo {volume} {973}},\ \bibinfo {pages} {L14} (\bibinfo {year} {2024})},\ \Eprint {https://arxiv.org/abs/2401.02929} {arXiv:2401.02929 [astro-ph.CO]} \BibitemShut {NoStop}%
\bibitem [{\citenamefont {Brout}\ \emph {et~al.}(2022)\citenamefont {Brout} \emph {et~al.}}]{Brout:2022vxf}%
  \BibitemOpen
  \bibfield  {author} {\bibinfo {author} {\bibfnamefont {D.}~\bibnamefont {Brout}} \emph {et~al.},\ }\bibfield  {title} {\bibinfo {title} {{The Pantheon+ Analysis: Cosmological Constraints}},\ }\href {https://doi.org/10.3847/1538-4357/ac8e04} {\bibfield  {journal} {\bibinfo  {journal} {Astrophys. J.}\ }\textbf {\bibinfo {volume} {938}},\ \bibinfo {pages} {110} (\bibinfo {year} {2022})},\ \Eprint {https://arxiv.org/abs/2202.04077} {arXiv:2202.04077 [astro-ph.CO]} \BibitemShut {NoStop}%
\bibitem [{\citenamefont {Rubin}\ \emph {et~al.}(2023)\citenamefont {Rubin} \emph {et~al.}}]{Rubin:2023jdq}%
  \BibitemOpen
  \bibfield  {author} {\bibinfo {author} {\bibfnamefont {D.}~\bibnamefont {Rubin}} \emph {et~al.},\ }\href@noop {} {\bibinfo {title} {{Union Through UNITY: Cosmology with 2,000 SNe Using a Unified Bayesian Framework}}} (\bibinfo {year} {2023}),\ \Eprint {https://arxiv.org/abs/2311.12098} {arXiv:2311.12098 [astro-ph.CO]} \BibitemShut {NoStop}%
\bibitem [{\citenamefont {Guth}(1981)}]{PhysRevD.23.347}%
  \BibitemOpen
  \bibfield  {author} {\bibinfo {author} {\bibfnamefont {A.~H.}\ \bibnamefont {Guth}},\ }\bibfield  {title} {\bibinfo {title} {Inflationary universe: A possible solution to the horizon and flatness problems},\ }\href {https://doi.org/10.1103/PhysRevD.23.347} {\bibfield  {journal} {\bibinfo  {journal} {Phys. Rev. D}\ }\textbf {\bibinfo {volume} {23}},\ \bibinfo {pages} {347} (\bibinfo {year} {1981})}\BibitemShut {NoStop}%
\bibitem [{\citenamefont {Starobinsky}(1980)}]{Starobinsky:1980te}%
  \BibitemOpen
  \bibfield  {author} {\bibinfo {author} {\bibfnamefont {A.~A.}\ \bibnamefont {Starobinsky}},\ }\bibfield  {title} {\bibinfo {title} {{A New Type of Isotropic Cosmological Models Without Singularity}},\ }\href {https://doi.org/10.1016/0370-2693(80)90670-X} {\bibfield  {journal} {\bibinfo  {journal} {Phys. Lett. B}\ }\textbf {\bibinfo {volume} {91}},\ \bibinfo {pages} {99} (\bibinfo {year} {1980})}\BibitemShut {NoStop}%
\bibitem [{\citenamefont {Sato}(1981)}]{Sato:1981qmu}%
  \BibitemOpen
  \bibfield  {author} {\bibinfo {author} {\bibfnamefont {K.}~\bibnamefont {Sato}},\ }\bibfield  {title} {\bibinfo {title} {{First-order phase transition of a vacuum and the expansion of the Universe}},\ }\href {https://doi.org/10.1093/mnras/195.3.467} {\bibfield  {journal} {\bibinfo  {journal} {Mon. Not. Roy. Astron. Soc.}\ }\textbf {\bibinfo {volume} {195}},\ \bibinfo {pages} {467} (\bibinfo {year} {1981})}\BibitemShut {NoStop}%
\bibitem [{\citenamefont {Linde}(1982)}]{Linde:1981mu}%
  \BibitemOpen
  \bibfield  {author} {\bibinfo {author} {\bibfnamefont {A.~D.}\ \bibnamefont {Linde}},\ }\bibfield  {title} {\bibinfo {title} {{A New Inflationary Universe Scenario: A Possible Solution of the Horizon, Flatness, Homogeneity, Isotropy and Primordial Monopole Problems}},\ }\href {https://doi.org/10.1016/0370-2693(82)91219-9} {\bibfield  {journal} {\bibinfo  {journal} {Phys. Lett. B}\ }\textbf {\bibinfo {volume} {108}},\ \bibinfo {pages} {389} (\bibinfo {year} {1982})}\BibitemShut {NoStop}%
\bibitem [{\citenamefont {Albrecht}\ and\ \citenamefont {Steinhardt}(1982)}]{Albrecht:1982wi}%
  \BibitemOpen
  \bibfield  {author} {\bibinfo {author} {\bibfnamefont {A.}~\bibnamefont {Albrecht}}\ and\ \bibinfo {author} {\bibfnamefont {P.~J.}\ \bibnamefont {Steinhardt}},\ }\bibfield  {title} {\bibinfo {title} {{Cosmology for Grand Unified Theories with Radiatively Induced Symmetry Breaking}},\ }\href {https://doi.org/10.1103/PhysRevLett.48.1220} {\bibfield  {journal} {\bibinfo  {journal} {Phys. Rev. Lett.}\ }\textbf {\bibinfo {volume} {48}},\ \bibinfo {pages} {1220} (\bibinfo {year} {1982})}\BibitemShut {NoStop}%
\bibitem [{\citenamefont {Mukhanov}\ \emph {et~al.}(1992)\citenamefont {Mukhanov}, \citenamefont {Feldman},\ and\ \citenamefont {Brandenberger}}]{Mukhanov:1990me}%
  \BibitemOpen
  \bibfield  {author} {\bibinfo {author} {\bibfnamefont {V.~F.}\ \bibnamefont {Mukhanov}}, \bibinfo {author} {\bibfnamefont {H.~A.}\ \bibnamefont {Feldman}},\ and\ \bibinfo {author} {\bibfnamefont {R.~H.}\ \bibnamefont {Brandenberger}},\ }\bibfield  {title} {\bibinfo {title} {{Theory of cosmological perturbations. Part 1. Classical perturbations. Part 2. Quantum theory of perturbations. Part 3. Extensions}},\ }\href {https://doi.org/10.1016/0370-1573(92)90044-Z} {\bibfield  {journal} {\bibinfo  {journal} {Phys. Rept.}\ }\textbf {\bibinfo {volume} {215}},\ \bibinfo {pages} {203} (\bibinfo {year} {1992})}\BibitemShut {NoStop}%
\bibitem [{\citenamefont {Starobinsky}(1979)}]{Starobinsky:1979ty}%
  \BibitemOpen
  \bibfield  {author} {\bibinfo {author} {\bibfnamefont {A.~A.}\ \bibnamefont {Starobinsky}},\ }\bibfield  {title} {\bibinfo {title} {{Spectrum of relict gravitational radiation and the early state of the universe}},\ }\href@noop {} {\bibfield  {journal} {\bibinfo  {journal} {JETP Lett.}\ }\textbf {\bibinfo {volume} {30}},\ \bibinfo {pages} {682} (\bibinfo {year} {1979})}\BibitemShut {NoStop}%
\bibitem [{\citenamefont {Di~Valentino}\ \emph {et~al.}(2021)\citenamefont {Di~Valentino}, \citenamefont {Mena}, \citenamefont {Pan}, \citenamefont {Visinelli}, \citenamefont {Yang}, \citenamefont {Melchiorri}, \citenamefont {Mota}, \citenamefont {Riess},\ and\ \citenamefont {Silk}}]{DiValentino:2021izs}%
  \BibitemOpen
  \bibfield  {author} {\bibinfo {author} {\bibfnamefont {E.}~\bibnamefont {Di~Valentino}}, \bibinfo {author} {\bibfnamefont {O.}~\bibnamefont {Mena}}, \bibinfo {author} {\bibfnamefont {S.}~\bibnamefont {Pan}}, \bibinfo {author} {\bibfnamefont {L.}~\bibnamefont {Visinelli}}, \bibinfo {author} {\bibfnamefont {W.}~\bibnamefont {Yang}}, \bibinfo {author} {\bibfnamefont {A.}~\bibnamefont {Melchiorri}}, \bibinfo {author} {\bibfnamefont {D.~F.}\ \bibnamefont {Mota}}, \bibinfo {author} {\bibfnamefont {A.~G.}\ \bibnamefont {Riess}},\ and\ \bibinfo {author} {\bibfnamefont {J.}~\bibnamefont {Silk}},\ }\bibfield  {title} {\bibinfo {title} {{In the realm of the Hubble tension{\textemdash}a review of solutions}},\ }\href {https://doi.org/10.1088/1361-6382/ac086d} {\bibfield  {journal} {\bibinfo  {journal} {Class. Quant. Grav.}\ }\textbf {\bibinfo {volume} {38}},\ \bibinfo {pages} {153001} (\bibinfo {year} {2021})},\ \Eprint {https://arxiv.org/abs/2103.01183} {arXiv:2103.01183 [astro-ph.CO]} \BibitemShut {NoStop}%
\bibitem [{\citenamefont {Chen}\ \emph {et~al.}(2024)\citenamefont {Chen}, \citenamefont {Ivanov}, \citenamefont {Philcox},\ and\ \citenamefont {Wenzl}}]{Chen:2024vuf}%
  \BibitemOpen
  \bibfield  {author} {\bibinfo {author} {\bibfnamefont {S.-F.}\ \bibnamefont {Chen}}, \bibinfo {author} {\bibfnamefont {M.~M.}\ \bibnamefont {Ivanov}}, \bibinfo {author} {\bibfnamefont {O.~H.~E.}\ \bibnamefont {Philcox}},\ and\ \bibinfo {author} {\bibfnamefont {L.}~\bibnamefont {Wenzl}},\ }\bibfield  {title} {\bibinfo {title} {{Suppression without Thawing: Constraining Structure Formation and Dark Energy with Galaxy Clustering}},\ }\href {https://doi.org/10.1103/PhysRevLett.133.231001} {\bibfield  {journal} {\bibinfo  {journal} {Phys. Rev. Lett.}\ }\textbf {\bibinfo {volume} {133}},\ \bibinfo {pages} {231001} (\bibinfo {year} {2024})},\ \Eprint {https://arxiv.org/abs/2406.13388} {arXiv:2406.13388 [astro-ph.CO]} \BibitemShut {NoStop}%
\bibitem [{\citenamefont {Lovelock}(1971)}]{Lovelock:1971yv}%
  \BibitemOpen
  \bibfield  {author} {\bibinfo {author} {\bibfnamefont {D.}~\bibnamefont {Lovelock}},\ }\bibfield  {title} {\bibinfo {title} {{The Einstein tensor and its generalizations}},\ }\href {https://doi.org/10.1063/1.1665613} {\bibfield  {journal} {\bibinfo  {journal} {J. Math. Phys.}\ }\textbf {\bibinfo {volume} {12}},\ \bibinfo {pages} {498} (\bibinfo {year} {1971})}\BibitemShut {NoStop}%
\bibitem [{\citenamefont {Heisenberg}(2019)}]{Heisenberg_2019}%
  \BibitemOpen
  \bibfield  {author} {\bibinfo {author} {\bibfnamefont {L.}~\bibnamefont {Heisenberg}},\ }\bibfield  {title} {\bibinfo {title} {A systematic approach to generalisations of general relativity and their cosmological implications},\ }\href {https://doi.org/10.1016/j.physrep.2018.11.006} {\bibfield  {journal} {\bibinfo  {journal} {Physics Reports}\ }\textbf {\bibinfo {volume} {796}},\ \bibinfo {pages} {1–113} (\bibinfo {year} {2019})}\BibitemShut {NoStop}%
\bibitem [{\citenamefont {Lan}\ \emph {et~al.}(2023)\citenamefont {Lan}, \citenamefont {Yang}, \citenamefont {Guo},\ and\ \citenamefont {Miao}}]{Lan:2023cvz}%
  \BibitemOpen
  \bibfield  {author} {\bibinfo {author} {\bibfnamefont {C.}~\bibnamefont {Lan}}, \bibinfo {author} {\bibfnamefont {H.}~\bibnamefont {Yang}}, \bibinfo {author} {\bibfnamefont {Y.}~\bibnamefont {Guo}},\ and\ \bibinfo {author} {\bibfnamefont {Y.-G.}\ \bibnamefont {Miao}},\ }\bibfield  {title} {\bibinfo {title} {{Regular Black Holes: A Short Topic Review}},\ }\href {https://doi.org/10.1007/s10773-023-05454-1} {\bibfield  {journal} {\bibinfo  {journal} {Int. J. Theor. Phys.}\ }\textbf {\bibinfo {volume} {62}},\ \bibinfo {pages} {202} (\bibinfo {year} {2023})},\ \Eprint {https://arxiv.org/abs/2303.11696} {arXiv:2303.11696 [gr-qc]} \BibitemShut {NoStop}%
\bibitem [{\citenamefont {Carballo-Rubio}\ \emph {et~al.}(2025)\citenamefont {Carballo-Rubio} \emph {et~al.}}]{Carballo-Rubio:2025fnc}%
  \BibitemOpen
  \bibfield  {author} {\bibinfo {author} {\bibfnamefont {R.}~\bibnamefont {Carballo-Rubio}} \emph {et~al.},\ }\href@noop {} {\bibinfo {title} {{Towards a Non-singular Paradigm of Black Hole Physics}}} (\bibinfo {year} {2025}),\ \Eprint {https://arxiv.org/abs/2501.05505} {arXiv:2501.05505 [gr-qc]} \BibitemShut {NoStop}%
\bibitem [{\citenamefont {Torres}(2022)}]{Torres:2022twv}%
  \BibitemOpen
  \bibfield  {author} {\bibinfo {author} {\bibfnamefont {R.}~\bibnamefont {Torres}},\ }\href@noop {} {\bibinfo {title} {{Regular Rotating Black Holes: A Review}}} (\bibinfo {year} {2022}),\ \Eprint {https://arxiv.org/abs/2208.12713} {arXiv:2208.12713 [gr-qc]} \BibitemShut {NoStop}%
\bibitem [{\citenamefont {{Bardeen}}(1968)}]{1968qtr..conf...87B}%
  \BibitemOpen
  \bibfield  {author} {\bibinfo {author} {\bibfnamefont {J.}~\bibnamefont {{Bardeen}}},\ }\bibfield  {title} {\bibinfo {title} {{Non-singular general relativistic gravitational collapse}},\ }in\ \href@noop {} {\emph {\bibinfo {booktitle} {Proceedings of the 5th International Conference on Gravitation and the Theory of Relativity}}}\ (\bibinfo {year} {1968})\ p.~\bibinfo {pages} {87}\BibitemShut {NoStop}%
\bibitem [{\citenamefont {Dymnikova}(1992)}]{Dymnikova:1992ux}%
  \BibitemOpen
  \bibfield  {author} {\bibinfo {author} {\bibfnamefont {I.}~\bibnamefont {Dymnikova}},\ }\bibfield  {title} {\bibinfo {title} {{Vacuum nonsingular black hole}},\ }\href {https://doi.org/10.1007/BF00760226} {\bibfield  {journal} {\bibinfo  {journal} {Gen. Rel. Grav.}\ }\textbf {\bibinfo {volume} {24}},\ \bibinfo {pages} {235} (\bibinfo {year} {1992})}\BibitemShut {NoStop}%
\bibitem [{\citenamefont {Hayward}(2006)}]{Hayward:2005gi}%
  \BibitemOpen
  \bibfield  {author} {\bibinfo {author} {\bibfnamefont {S.~A.}\ \bibnamefont {Hayward}},\ }\bibfield  {title} {\bibinfo {title} {{Formation and evaporation of regular black holes}},\ }\href {https://doi.org/10.1103/PhysRevLett.96.031103} {\bibfield  {journal} {\bibinfo  {journal} {Phys. Rev. Lett.}\ }\textbf {\bibinfo {volume} {96}},\ \bibinfo {pages} {031103} (\bibinfo {year} {2006})},\ \Eprint {https://arxiv.org/abs/gr-qc/0506126} {arXiv:gr-qc/0506126} \BibitemShut {NoStop}%
\bibitem [{\citenamefont {Bambi}\ and\ \citenamefont {Modesto}(2013)}]{Bambi:2013ufa}%
  \BibitemOpen
  \bibfield  {author} {\bibinfo {author} {\bibfnamefont {C.}~\bibnamefont {Bambi}}\ and\ \bibinfo {author} {\bibfnamefont {L.}~\bibnamefont {Modesto}},\ }\bibfield  {title} {\bibinfo {title} {{Rotating regular black holes}},\ }\href {https://doi.org/10.1016/j.physletb.2013.03.025} {\bibfield  {journal} {\bibinfo  {journal} {Phys. Lett. B}\ }\textbf {\bibinfo {volume} {721}},\ \bibinfo {pages} {329} (\bibinfo {year} {2013})},\ \Eprint {https://arxiv.org/abs/1302.6075} {arXiv:1302.6075 [gr-qc]} \BibitemShut {NoStop}%
\bibitem [{\citenamefont {Simpson}\ and\ \citenamefont {Visser}(2019)}]{Simpson:2019mud}%
  \BibitemOpen
  \bibfield  {author} {\bibinfo {author} {\bibfnamefont {A.}~\bibnamefont {Simpson}}\ and\ \bibinfo {author} {\bibfnamefont {M.}~\bibnamefont {Visser}},\ }\bibfield  {title} {\bibinfo {title} {{Regular black holes with asymptotically Minkowski cores}},\ }\href {https://doi.org/10.3390/universe6010008} {\bibfield  {journal} {\bibinfo  {journal} {Universe}\ }\textbf {\bibinfo {volume} {6}},\ \bibinfo {pages} {8} (\bibinfo {year} {2019})},\ \Eprint {https://arxiv.org/abs/1911.01020} {arXiv:1911.01020 [gr-qc]} \BibitemShut {NoStop}%
\bibitem [{\citenamefont {Carballo-Rubio}\ \emph {et~al.}(2023{\natexlab{a}})\citenamefont {Carballo-Rubio}, \citenamefont {Di~Filippo}, \citenamefont {Liberati},\ and\ \citenamefont {Visser}}]{Carballo-Rubio:2023mvr}%
  \BibitemOpen
  \bibfield  {author} {\bibinfo {author} {\bibfnamefont {R.}~\bibnamefont {Carballo-Rubio}}, \bibinfo {author} {\bibfnamefont {F.}~\bibnamefont {Di~Filippo}}, \bibinfo {author} {\bibfnamefont {S.}~\bibnamefont {Liberati}},\ and\ \bibinfo {author} {\bibfnamefont {M.}~\bibnamefont {Visser}},\ }\href@noop {} {\bibinfo {title} {{Singularity-free gravitational collapse: From regular black holes to horizonless objects}}} (\bibinfo {year} {2023}{\natexlab{a}}),\ \Eprint {https://arxiv.org/abs/2302.00028} {arXiv:2302.00028 [gr-qc]} \BibitemShut {NoStop}%
\bibitem [{\citenamefont {Carballo-Rubio}\ \emph {et~al.}(2023{\natexlab{b}})\citenamefont {Carballo-Rubio}, \citenamefont {Di~Filippo}, \citenamefont {Liberati},\ and\ \citenamefont {Visser}}]{Carballo-Rubio:2022nuj}%
  \BibitemOpen
  \bibfield  {author} {\bibinfo {author} {\bibfnamefont {R.}~\bibnamefont {Carballo-Rubio}}, \bibinfo {author} {\bibfnamefont {F.}~\bibnamefont {Di~Filippo}}, \bibinfo {author} {\bibfnamefont {S.}~\bibnamefont {Liberati}},\ and\ \bibinfo {author} {\bibfnamefont {M.}~\bibnamefont {Visser}},\ }\bibfield  {title} {\bibinfo {title} {{A connection between regular black holes and horizonless ultracompact stars}},\ }\href {https://doi.org/10.1007/JHEP08(2023)046} {\bibfield  {journal} {\bibinfo  {journal} {JHEP}\ }\textbf {\bibinfo {volume} {08}},\ \bibinfo {pages} {046}},\ \Eprint {https://arxiv.org/abs/2211.05817} {arXiv:2211.05817 [gr-qc]} \BibitemShut {NoStop}%
\bibitem [{\citenamefont {Carballo-Rubio}\ \emph {et~al.}(2022)\citenamefont {Carballo-Rubio}, \citenamefont {Di~Filippo}, \citenamefont {Liberati}, \citenamefont {Pacilio},\ and\ \citenamefont {Visser}}]{Carballo-Rubio:2022kad}%
  \BibitemOpen
  \bibfield  {author} {\bibinfo {author} {\bibfnamefont {R.}~\bibnamefont {Carballo-Rubio}}, \bibinfo {author} {\bibfnamefont {F.}~\bibnamefont {Di~Filippo}}, \bibinfo {author} {\bibfnamefont {S.}~\bibnamefont {Liberati}}, \bibinfo {author} {\bibfnamefont {C.}~\bibnamefont {Pacilio}},\ and\ \bibinfo {author} {\bibfnamefont {M.}~\bibnamefont {Visser}},\ }\bibfield  {title} {\bibinfo {title} {{Regular black holes without mass inflation instability}},\ }\href {https://doi.org/10.1007/JHEP09(2022)118} {\bibfield  {journal} {\bibinfo  {journal} {JHEP}\ }\textbf {\bibinfo {volume} {09}},\ \bibinfo {pages} {118}},\ \Eprint {https://arxiv.org/abs/2205.13556} {arXiv:2205.13556 [gr-qc]} \BibitemShut {NoStop}%
\bibitem [{\citenamefont {Di~Filippo}\ \emph {et~al.}(2022)\citenamefont {Di~Filippo}, \citenamefont {Carballo-Rubio}, \citenamefont {Liberati}, \citenamefont {Pacilio},\ and\ \citenamefont {Visser}}]{DiFilippo:2022qkl}%
  \BibitemOpen
  \bibfield  {author} {\bibinfo {author} {\bibfnamefont {F.}~\bibnamefont {Di~Filippo}}, \bibinfo {author} {\bibfnamefont {R.}~\bibnamefont {Carballo-Rubio}}, \bibinfo {author} {\bibfnamefont {S.}~\bibnamefont {Liberati}}, \bibinfo {author} {\bibfnamefont {C.}~\bibnamefont {Pacilio}},\ and\ \bibinfo {author} {\bibfnamefont {M.}~\bibnamefont {Visser}},\ }\bibfield  {title} {\bibinfo {title} {{On the Inner Horizon Instability of Non-Singular Black Holes}},\ }\href {https://doi.org/10.3390/universe8040204} {\bibfield  {journal} {\bibinfo  {journal} {Universe}\ }\textbf {\bibinfo {volume} {8}},\ \bibinfo {pages} {204} (\bibinfo {year} {2022})},\ \Eprint {https://arxiv.org/abs/2203.14516} {arXiv:2203.14516 [gr-qc]} \BibitemShut {NoStop}%
\bibitem [{\citenamefont {Carballo-Rubio}\ \emph {et~al.}(2021)\citenamefont {Carballo-Rubio}, \citenamefont {Di~Filippo}, \citenamefont {Liberati}, \citenamefont {Pacilio},\ and\ \citenamefont {Visser}}]{Carballo-Rubio:2021bpr}%
  \BibitemOpen
  \bibfield  {author} {\bibinfo {author} {\bibfnamefont {R.}~\bibnamefont {Carballo-Rubio}}, \bibinfo {author} {\bibfnamefont {F.}~\bibnamefont {Di~Filippo}}, \bibinfo {author} {\bibfnamefont {S.}~\bibnamefont {Liberati}}, \bibinfo {author} {\bibfnamefont {C.}~\bibnamefont {Pacilio}},\ and\ \bibinfo {author} {\bibfnamefont {M.}~\bibnamefont {Visser}},\ }\bibfield  {title} {\bibinfo {title} {{Inner horizon instability and the unstable cores of regular black holes}},\ }\href {https://doi.org/10.1007/JHEP05(2021)132} {\bibfield  {journal} {\bibinfo  {journal} {JHEP}\ }\textbf {\bibinfo {volume} {05}},\ \bibinfo {pages} {132}},\ \Eprint {https://arxiv.org/abs/2101.05006} {arXiv:2101.05006 [gr-qc]} \BibitemShut {NoStop}%
\bibitem [{\citenamefont {Carballo-Rubio}\ \emph {et~al.}(2020)\citenamefont {Carballo-Rubio}, \citenamefont {Di~Filippo}, \citenamefont {Liberati},\ and\ \citenamefont {Visser}}]{Carballo-Rubio:2019fnb}%
  \BibitemOpen
  \bibfield  {author} {\bibinfo {author} {\bibfnamefont {R.}~\bibnamefont {Carballo-Rubio}}, \bibinfo {author} {\bibfnamefont {F.}~\bibnamefont {Di~Filippo}}, \bibinfo {author} {\bibfnamefont {S.}~\bibnamefont {Liberati}},\ and\ \bibinfo {author} {\bibfnamefont {M.}~\bibnamefont {Visser}},\ }\bibfield  {title} {\bibinfo {title} {{Geodesically complete black holes}},\ }\href {https://doi.org/10.1103/PhysRevD.101.084047} {\bibfield  {journal} {\bibinfo  {journal} {Phys. Rev. D}\ }\textbf {\bibinfo {volume} {101}},\ \bibinfo {pages} {084047} (\bibinfo {year} {2020})},\ \Eprint {https://arxiv.org/abs/1911.11200} {arXiv:1911.11200 [gr-qc]} \BibitemShut {NoStop}%
\bibitem [{\citenamefont {Carballo-Rubio}\ \emph {et~al.}(2018)\citenamefont {Carballo-Rubio}, \citenamefont {Di~Filippo}, \citenamefont {Liberati}, \citenamefont {Pacilio},\ and\ \citenamefont {Visser}}]{Carballo-Rubio:2018pmi}%
  \BibitemOpen
  \bibfield  {author} {\bibinfo {author} {\bibfnamefont {R.}~\bibnamefont {Carballo-Rubio}}, \bibinfo {author} {\bibfnamefont {F.}~\bibnamefont {Di~Filippo}}, \bibinfo {author} {\bibfnamefont {S.}~\bibnamefont {Liberati}}, \bibinfo {author} {\bibfnamefont {C.}~\bibnamefont {Pacilio}},\ and\ \bibinfo {author} {\bibfnamefont {M.}~\bibnamefont {Visser}},\ }\bibfield  {title} {\bibinfo {title} {{On the viability of regular black holes}},\ }\href {https://doi.org/10.1007/JHEP07(2018)023} {\bibfield  {journal} {\bibinfo  {journal} {JHEP}\ }\textbf {\bibinfo {volume} {07}},\ \bibinfo {pages} {023}},\ \Eprint {https://arxiv.org/abs/1805.02675} {arXiv:1805.02675 [gr-qc]} \BibitemShut {NoStop}%
\bibitem [{\citenamefont {Borissova}\ \emph {et~al.}(2025)\citenamefont {Borissova}, \citenamefont {Liberati},\ and\ \citenamefont {Visser}}]{Borissova:2025msp}%
  \BibitemOpen
  \bibfield  {author} {\bibinfo {author} {\bibfnamefont {J.}~\bibnamefont {Borissova}}, \bibinfo {author} {\bibfnamefont {S.}~\bibnamefont {Liberati}},\ and\ \bibinfo {author} {\bibfnamefont {M.}~\bibnamefont {Visser}},\ }\bibfield  {title} {\bibinfo {title} {{Violations of the null convergence condition in kinematical transitions between singular and regular black holes, horizonless compact objects, and bounces}},\ }\href {https://doi.org/10.1103/PhysRevD.111.104054} {\bibfield  {journal} {\bibinfo  {journal} {Phys. Rev. D}\ }\textbf {\bibinfo {volume} {111}},\ \bibinfo {pages} {104054} (\bibinfo {year} {2025})},\ \Eprint {https://arxiv.org/abs/2502.00548} {arXiv:2502.00548 [gr-qc]} \BibitemShut {NoStop}%
\bibitem [{\citenamefont {Di~Filippo}\ \emph {et~al.}(2024)\citenamefont {Di~Filippo}, \citenamefont {Liberati},\ and\ \citenamefont {Visser}}]{DiFilippo:2024spj}%
  \BibitemOpen
  \bibfield  {author} {\bibinfo {author} {\bibfnamefont {F.}~\bibnamefont {Di~Filippo}}, \bibinfo {author} {\bibfnamefont {S.}~\bibnamefont {Liberati}},\ and\ \bibinfo {author} {\bibfnamefont {M.}~\bibnamefont {Visser}},\ }\bibfield  {title} {\bibinfo {title} {{Fully extremal black holes: A black hole graveyard?}},\ }\href {https://doi.org/10.1142/S0218271824400054} {\bibfield  {journal} {\bibinfo  {journal} {Int. J. Mod. Phys. D}\ }\textbf {\bibinfo {volume} {33}},\ \bibinfo {pages} {2440005} (\bibinfo {year} {2024})},\ \Eprint {https://arxiv.org/abs/2405.08069} {arXiv:2405.08069 [gr-qc]} \BibitemShut {NoStop}%
\bibitem [{\citenamefont {Simpson}\ and\ \citenamefont {Visser}(2022)}]{Simpson:2021dyo}%
  \BibitemOpen
  \bibfield  {author} {\bibinfo {author} {\bibfnamefont {A.}~\bibnamefont {Simpson}}\ and\ \bibinfo {author} {\bibfnamefont {M.}~\bibnamefont {Visser}},\ }\bibfield  {title} {\bibinfo {title} {{The eye of the storm: a regular Kerr black hole}},\ }\href {https://doi.org/10.1088/1475-7516/2022/03/011} {\bibfield  {journal} {\bibinfo  {journal} {JCAP}\ }\textbf {\bibinfo {volume} {03}}\bibfield  {number} {\bibinfo  {number} { (03)},\ \bibinfo {pages} {011}},\ }\Eprint {https://arxiv.org/abs/2111.12329} {arXiv:2111.12329 [gr-qc]} \BibitemShut {NoStop}%
\bibitem [{\citenamefont {Arrechea}\ \emph {et~al.}(2025)\citenamefont {Arrechea}, \citenamefont {Liberati}, \citenamefont {Neshat},\ and\ \citenamefont {Vellucci}}]{Arrechea:2025nlq}%
  \BibitemOpen
  \bibfield  {author} {\bibinfo {author} {\bibfnamefont {J.}~\bibnamefont {Arrechea}}, \bibinfo {author} {\bibfnamefont {S.}~\bibnamefont {Liberati}}, \bibinfo {author} {\bibfnamefont {H.}~\bibnamefont {Neshat}},\ and\ \bibinfo {author} {\bibfnamefont {V.}~\bibnamefont {Vellucci}},\ }\href@noop {} {\bibinfo {title} {{From de Sitter to Anti-de Sitter singularity regularization: theory and phenomenology}}} (\bibinfo {year} {2025}),\ \Eprint {https://arxiv.org/abs/2509.13421} {arXiv:2509.13421 [gr-qc]} \BibitemShut {NoStop}%
\bibitem [{\citenamefont {Coviello}\ \emph {et~al.}(2025)\citenamefont {Coviello}, \citenamefont {Vellucci},\ and\ \citenamefont {Lehner}}]{Coviello:2025pla}%
  \BibitemOpen
  \bibfield  {author} {\bibinfo {author} {\bibfnamefont {C.}~\bibnamefont {Coviello}}, \bibinfo {author} {\bibfnamefont {V.}~\bibnamefont {Vellucci}},\ and\ \bibinfo {author} {\bibfnamefont {L.}~\bibnamefont {Lehner}},\ }\bibfield  {title} {\bibinfo {title} {{Tidal response of regular black holes}},\ }\href {https://doi.org/10.1103/PhysRevD.111.104073} {\bibfield  {journal} {\bibinfo  {journal} {Phys. Rev. D}\ }\textbf {\bibinfo {volume} {111}},\ \bibinfo {pages} {104073} (\bibinfo {year} {2025})},\ \Eprint {https://arxiv.org/abs/2503.04287} {arXiv:2503.04287 [gr-qc]} \BibitemShut {NoStop}%
\bibitem [{\citenamefont {Franzin}\ \emph {et~al.}(2024)\citenamefont {Franzin}, \citenamefont {Liberati},\ and\ \citenamefont {Vellucci}}]{Franzin:2023slm}%
  \BibitemOpen
  \bibfield  {author} {\bibinfo {author} {\bibfnamefont {E.}~\bibnamefont {Franzin}}, \bibinfo {author} {\bibfnamefont {S.}~\bibnamefont {Liberati}},\ and\ \bibinfo {author} {\bibfnamefont {V.}~\bibnamefont {Vellucci}},\ }\bibfield  {title} {\bibinfo {title} {{From regular black holes to horizonless objects: quasi-normal modes, instabilities and spectroscopy}},\ }\href {https://doi.org/10.1088/1475-7516/2024/01/020} {\bibfield  {journal} {\bibinfo  {journal} {JCAP}\ }\textbf {\bibinfo {volume} {01}},\ \bibinfo {pages} {020}},\ \Eprint {https://arxiv.org/abs/2310.11990} {arXiv:2310.11990 [gr-qc]} \BibitemShut {NoStop}%
\bibitem [{\citenamefont {Franzin}\ \emph {et~al.}(2022)\citenamefont {Franzin}, \citenamefont {Liberati}, \citenamefont {Mazza},\ and\ \citenamefont {Vellucci}}]{Franzin:2022wai}%
  \BibitemOpen
  \bibfield  {author} {\bibinfo {author} {\bibfnamefont {E.}~\bibnamefont {Franzin}}, \bibinfo {author} {\bibfnamefont {S.}~\bibnamefont {Liberati}}, \bibinfo {author} {\bibfnamefont {J.}~\bibnamefont {Mazza}},\ and\ \bibinfo {author} {\bibfnamefont {V.}~\bibnamefont {Vellucci}},\ }\bibfield  {title} {\bibinfo {title} {{Stable rotating regular black holes}},\ }\href {https://doi.org/10.1103/PhysRevD.106.104060} {\bibfield  {journal} {\bibinfo  {journal} {Phys. Rev. D}\ }\textbf {\bibinfo {volume} {106}},\ \bibinfo {pages} {104060} (\bibinfo {year} {2022})},\ \Eprint {https://arxiv.org/abs/2207.08864} {arXiv:2207.08864 [gr-qc]} \BibitemShut {NoStop}%
\bibitem [{\citenamefont {{Aguilar-Gutierrez}}\ \emph {et~al.}(2023)\citenamefont {{Aguilar-Gutierrez}}, \citenamefont {Bueno}, \citenamefont {Cano}, \citenamefont {Hennigar},\ and\ \citenamefont {Llorens}}]{aguilar-gutierrezAspectsHighercurvatureGravities2023}%
  \BibitemOpen
  \bibfield  {author} {\bibinfo {author} {\bibfnamefont {S.~E.}\ \bibnamefont {{Aguilar-Gutierrez}}}, \bibinfo {author} {\bibfnamefont {P.}~\bibnamefont {Bueno}}, \bibinfo {author} {\bibfnamefont {P.~A.}\ \bibnamefont {Cano}}, \bibinfo {author} {\bibfnamefont {R.~A.}\ \bibnamefont {Hennigar}},\ and\ \bibinfo {author} {\bibfnamefont {Q.}~\bibnamefont {Llorens}},\ }\href {https://doi.org/10.48550/arXiv.2310.09333} {\bibinfo {title} {Aspects of higher-curvature gravities with covariant derivatives}} (\bibinfo {year} {2023}),\ \Eprint {https://arxiv.org/abs/2310.09333} {arXiv:2310.09333} \BibitemShut {NoStop}%
\bibitem [{\citenamefont {Ahmed}\ \emph {et~al.}(2017)\citenamefont {Ahmed}, \citenamefont {Hennigar}, \citenamefont {Mann},\ and\ \citenamefont {Mir}}]{ahmedQuintessentialQuarticQuasitopological2017}%
  \BibitemOpen
  \bibfield  {author} {\bibinfo {author} {\bibfnamefont {J.}~\bibnamefont {Ahmed}}, \bibinfo {author} {\bibfnamefont {R.~A.}\ \bibnamefont {Hennigar}}, \bibinfo {author} {\bibfnamefont {R.~B.}\ \bibnamefont {Mann}},\ and\ \bibinfo {author} {\bibfnamefont {M.}~\bibnamefont {Mir}},\ }\href {https://doi.org/10.48550/arXiv.1703.11007} {\bibinfo {title} {Quintessential {{Quartic Quasi-topological Quartet}}}} (\bibinfo {year} {2017}),\ \Eprint {https://arxiv.org/abs/1703.11007} {arXiv:1703.11007} \BibitemShut {NoStop}%
\bibitem [{\citenamefont {Bueno}\ and\ \citenamefont {Cano}(2017{\natexlab{a}})}]{buenoFourdimensionalBlackHoles2017}%
  \BibitemOpen
  \bibfield  {author} {\bibinfo {author} {\bibfnamefont {P.}~\bibnamefont {Bueno}}\ and\ \bibinfo {author} {\bibfnamefont {P.~A.}\ \bibnamefont {Cano}},\ }\href {https://doi.org/10.48550/arXiv.1610.08019} {\bibinfo {title} {Four-dimensional black holes in {{Einsteinian}} cubic gravity}} (\bibinfo {year} {2017}{\natexlab{a}}),\ \Eprint {https://arxiv.org/abs/1610.08019} {arXiv:1610.08019} \BibitemShut {NoStop}%
\bibitem [{\citenamefont {Bueno}\ \emph {et~al.}(2019)\citenamefont {Bueno}, \citenamefont {Cano},\ and\ \citenamefont {Hennigar}}]{buenoGeneralizedQuasitopologicalGravities2019}%
  \BibitemOpen
  \bibfield  {author} {\bibinfo {author} {\bibfnamefont {P.}~\bibnamefont {Bueno}}, \bibinfo {author} {\bibfnamefont {P.~A.}\ \bibnamefont {Cano}},\ and\ \bibinfo {author} {\bibfnamefont {R.~A.}\ \bibnamefont {Hennigar}},\ }\href {https://doi.org/10.48550/arXiv.1909.07983} {\bibinfo {title} {({{Generalized}}) quasi-topological gravities at all orders}} (\bibinfo {year} {2019}),\ \Eprint {https://arxiv.org/abs/1909.07983} {arXiv:1909.07983} \BibitemShut {NoStop}%
\bibitem [{\citenamefont {Bueno}\ \emph {et~al.}(2022)\citenamefont {Bueno}, \citenamefont {Cano}, \citenamefont {Hennigar}, \citenamefont {Lu},\ and\ \citenamefont {Moreno}}]{buenoGeneralizedQuasitopologicalGravities2022}%
  \BibitemOpen
  \bibfield  {author} {\bibinfo {author} {\bibfnamefont {P.}~\bibnamefont {Bueno}}, \bibinfo {author} {\bibfnamefont {P.~A.}\ \bibnamefont {Cano}}, \bibinfo {author} {\bibfnamefont {R.~A.}\ \bibnamefont {Hennigar}}, \bibinfo {author} {\bibfnamefont {M.}~\bibnamefont {Lu}},\ and\ \bibinfo {author} {\bibfnamefont {J.}~\bibnamefont {Moreno}},\ }\href {https://doi.org/10.48550/arXiv.2203.05589} {\bibinfo {title} {Generalized quasi-topological gravities: The whole shebang}} (\bibinfo {year} {2022}),\ \Eprint {https://arxiv.org/abs/2203.05589} {arXiv:2203.05589} \BibitemShut {NoStop}%
\bibitem [{\citenamefont {Bueno}\ and\ \citenamefont {Cano}(2017{\natexlab{b}})}]{buenoUniversalBlackHole2017}%
  \BibitemOpen
  \bibfield  {author} {\bibinfo {author} {\bibfnamefont {P.}~\bibnamefont {Bueno}}\ and\ \bibinfo {author} {\bibfnamefont {P.~A.}\ \bibnamefont {Cano}},\ }\href {https://doi.org/10.48550/arXiv.1704.02967} {\bibinfo {title} {Universal black hole stability in four dimensions}} (\bibinfo {year} {2017}{\natexlab{b}}),\ \Eprint {https://arxiv.org/abs/1704.02967} {arXiv:1704.02967} \BibitemShut {NoStop}%
\bibitem [{\citenamefont {Hennigar}\ and\ \citenamefont {Mann}(2016)}]{hennigarBlackHolesEinsteinian2016}%
  \BibitemOpen
  \bibfield  {author} {\bibinfo {author} {\bibfnamefont {R.~A.}\ \bibnamefont {Hennigar}}\ and\ \bibinfo {author} {\bibfnamefont {R.~B.}\ \bibnamefont {Mann}},\ }\href {https://doi.org/10.48550/arXiv.1610.06675} {\bibinfo {title} {Black holes in {{Einsteinian}} cubic gravity}} (\bibinfo {year} {2016}),\ \Eprint {https://arxiv.org/abs/1610.06675} {arXiv:1610.06675} \BibitemShut {NoStop}%
\bibitem [{\citenamefont {Hennigar}\ \emph {et~al.}(2017)\citenamefont {Hennigar}, \citenamefont {Kubiznak},\ and\ \citenamefont {Mann}}]{hennigarGeneralizedQuasitopologicalGravity2017}%
  \BibitemOpen
  \bibfield  {author} {\bibinfo {author} {\bibfnamefont {R.~A.}\ \bibnamefont {Hennigar}}, \bibinfo {author} {\bibfnamefont {D.}~\bibnamefont {Kubiznak}},\ and\ \bibinfo {author} {\bibfnamefont {R.~B.}\ \bibnamefont {Mann}},\ }\href {https://doi.org/10.48550/arXiv.1703.01631} {\bibinfo {title} {Generalized quasi-topological gravity}} (\bibinfo {year} {2017}),\ \Eprint {https://arxiv.org/abs/1703.01631} {arXiv:1703.01631} \BibitemShut {NoStop}%
\bibitem [{\citenamefont {Moreno}\ and\ \citenamefont {Murcia}(2023)}]{morenoClassificationGeneralizedQuasitopological2023a}%
  \BibitemOpen
  \bibfield  {author} {\bibinfo {author} {\bibfnamefont {J.}~\bibnamefont {Moreno}}\ and\ \bibinfo {author} {\bibfnamefont {{\'A}.~J.}\ \bibnamefont {Murcia}},\ }\href {https://doi.org/10.48550/arXiv.2304.08510} {\bibinfo {title} {On the classification of {{Generalized Quasitopological Gravities}}}} (\bibinfo {year} {2023}),\ \Eprint {https://arxiv.org/abs/2304.08510} {arXiv:2304.08510} \BibitemShut {NoStop}%
\bibitem [{\citenamefont {Moreno}\ and\ \citenamefont {Murcia}(2024)}]{Moreno:2023arp}%
  \BibitemOpen
  \bibfield  {author} {\bibinfo {author} {\bibfnamefont {J.}~\bibnamefont {Moreno}}\ and\ \bibinfo {author} {\bibfnamefont {A.~J.}\ \bibnamefont {Murcia}},\ }\bibfield  {title} {\bibinfo {title} {{Cosmological higher-curvature gravities}},\ }\href {https://doi.org/10.1088/1361-6382/ad51c5} {\bibfield  {journal} {\bibinfo  {journal} {Class. Quant. Grav.}\ }\textbf {\bibinfo {volume} {41}},\ \bibinfo {pages} {135017} (\bibinfo {year} {2024})},\ \Eprint {https://arxiv.org/abs/2311.12104} {arXiv:2311.12104 [gr-qc]} \BibitemShut {NoStop}%
\bibitem [{\citenamefont {Ayon-Beato}\ and\ \citenamefont {Garcia}(1998)}]{Ayon-Beato:1998hmi}%
  \BibitemOpen
  \bibfield  {author} {\bibinfo {author} {\bibfnamefont {E.}~\bibnamefont {Ayon-Beato}}\ and\ \bibinfo {author} {\bibfnamefont {A.}~\bibnamefont {Garcia}},\ }\bibfield  {title} {\bibinfo {title} {{Regular black hole in general relativity coupled to nonlinear electrodynamics}},\ }\href {https://doi.org/10.1103/PhysRevLett.80.5056} {\bibfield  {journal} {\bibinfo  {journal} {Phys. Rev. Lett.}\ }\textbf {\bibinfo {volume} {80}},\ \bibinfo {pages} {5056} (\bibinfo {year} {1998})},\ \Eprint {https://arxiv.org/abs/gr-qc/9911046} {arXiv:gr-qc/9911046} \BibitemShut {NoStop}%
\bibitem [{\citenamefont {Bronnikov}(2001)}]{Bronnikov:2000vy}%
  \BibitemOpen
  \bibfield  {author} {\bibinfo {author} {\bibfnamefont {K.~A.}\ \bibnamefont {Bronnikov}},\ }\bibfield  {title} {\bibinfo {title} {{Regular magnetic black holes and monopoles from nonlinear electrodynamics}},\ }\href {https://doi.org/10.1103/PhysRevD.63.044005} {\bibfield  {journal} {\bibinfo  {journal} {Phys. Rev. D}\ }\textbf {\bibinfo {volume} {63}},\ \bibinfo {pages} {044005} (\bibinfo {year} {2001})},\ \Eprint {https://arxiv.org/abs/gr-qc/0006014} {arXiv:gr-qc/0006014} \BibitemShut {NoStop}%
\bibitem [{\citenamefont {Ayon-Beato}\ and\ \citenamefont {Garcia}(2000)}]{Ayon-Beato:2000mjt}%
  \BibitemOpen
  \bibfield  {author} {\bibinfo {author} {\bibfnamefont {E.}~\bibnamefont {Ayon-Beato}}\ and\ \bibinfo {author} {\bibfnamefont {A.}~\bibnamefont {Garcia}},\ }\bibfield  {title} {\bibinfo {title} {{The Bardeen model as a nonlinear magnetic monopole}},\ }\href {https://doi.org/10.1016/S0370-2693(00)01125-4} {\bibfield  {journal} {\bibinfo  {journal} {Phys. Lett. B}\ }\textbf {\bibinfo {volume} {493}},\ \bibinfo {pages} {149} (\bibinfo {year} {2000})},\ \Eprint {https://arxiv.org/abs/gr-qc/0009077} {arXiv:gr-qc/0009077} \BibitemShut {NoStop}%
\bibitem [{\citenamefont {Dymnikova}(2004)}]{Dymnikova:2004zc}%
  \BibitemOpen
  \bibfield  {author} {\bibinfo {author} {\bibfnamefont {I.}~\bibnamefont {Dymnikova}},\ }\bibfield  {title} {\bibinfo {title} {{Regular electrically charged structures in nonlinear electrodynamics coupled to general relativity}},\ }\href {https://doi.org/10.1088/0264-9381/21/18/009} {\bibfield  {journal} {\bibinfo  {journal} {Class. Quant. Grav.}\ }\textbf {\bibinfo {volume} {21}},\ \bibinfo {pages} {4417} (\bibinfo {year} {2004})},\ \Eprint {https://arxiv.org/abs/gr-qc/0407072} {arXiv:gr-qc/0407072} \BibitemShut {NoStop}%
\bibitem [{\citenamefont {Bronnikov}(2018)}]{Bronnikov:2017sgg}%
  \BibitemOpen
  \bibfield  {author} {\bibinfo {author} {\bibfnamefont {K.~A.}\ \bibnamefont {Bronnikov}},\ }\bibfield  {title} {\bibinfo {title} {{Nonlinear electrodynamics, regular black holes and wormholes}},\ }\href {https://doi.org/10.1142/S0218271818410055} {\bibfield  {journal} {\bibinfo  {journal} {Int. J. Mod. Phys. D}\ }\textbf {\bibinfo {volume} {27}},\ \bibinfo {pages} {1841005} (\bibinfo {year} {2018})},\ \Eprint {https://arxiv.org/abs/1711.00087} {arXiv:1711.00087 [gr-qc]} \BibitemShut {NoStop}%
\bibitem [{\citenamefont {Cano}\ and\ \citenamefont {Murcia}(2020)}]{Cano:2020qhy}%
  \BibitemOpen
  \bibfield  {author} {\bibinfo {author} {\bibfnamefont {P.~A.}\ \bibnamefont {Cano}}\ and\ \bibinfo {author} {\bibfnamefont {A.}~\bibnamefont {Murcia}},\ }\bibfield  {title} {\bibinfo {title} {{Electromagnetic Quasitopological Gravities}},\ }\href {https://doi.org/10.1007/JHEP10(2020)125} {\bibfield  {journal} {\bibinfo  {journal} {JHEP}\ }\textbf {\bibinfo {volume} {10}},\ \bibinfo {pages} {125}},\ \Eprint {https://arxiv.org/abs/2007.04331} {arXiv:2007.04331 [hep-th]} \BibitemShut {NoStop}%
\bibitem [{\citenamefont {Cano}\ and\ \citenamefont {Murcia}(2021)}]{Cano:2020ezi}%
  \BibitemOpen
  \bibfield  {author} {\bibinfo {author} {\bibfnamefont {P.~A.}\ \bibnamefont {Cano}}\ and\ \bibinfo {author} {\bibfnamefont {A.}~\bibnamefont {Murcia}},\ }\bibfield  {title} {\bibinfo {title} {{Resolution of Reissner-Nordstr\"om singularities by higher-derivative corrections}},\ }\href {https://doi.org/10.1088/1361-6382/abd923} {\bibfield  {journal} {\bibinfo  {journal} {Class. Quant. Grav.}\ }\textbf {\bibinfo {volume} {38}},\ \bibinfo {pages} {075014} (\bibinfo {year} {2021})},\ \Eprint {https://arxiv.org/abs/2006.15149} {arXiv:2006.15149 [hep-th]} \BibitemShut {NoStop}%
\bibitem [{\citenamefont {Babichev}\ \emph {et~al.}(2020)\citenamefont {Babichev}, \citenamefont {Charmousis}, \citenamefont {Cisterna},\ and\ \citenamefont {Hassaine}}]{Babichev:2020qpr}%
  \BibitemOpen
  \bibfield  {author} {\bibinfo {author} {\bibfnamefont {E.}~\bibnamefont {Babichev}}, \bibinfo {author} {\bibfnamefont {C.}~\bibnamefont {Charmousis}}, \bibinfo {author} {\bibfnamefont {A.}~\bibnamefont {Cisterna}},\ and\ \bibinfo {author} {\bibfnamefont {M.}~\bibnamefont {Hassaine}},\ }\bibfield  {title} {\bibinfo {title} {{Regular black holes via the Kerr-Schild construction in DHOST theories}},\ }\href {https://doi.org/10.1088/1475-7516/2020/06/049} {\bibfield  {journal} {\bibinfo  {journal} {JCAP}\ }\textbf {\bibinfo {volume} {06}},\ \bibinfo {pages} {049}},\ \Eprint {https://arxiv.org/abs/2004.00597} {arXiv:2004.00597 [hep-th]} \BibitemShut {NoStop}%
\bibitem [{\citenamefont {Baake}\ \emph {et~al.}(2021)\citenamefont {Baake}, \citenamefont {Charmousis}, \citenamefont {Hassaine},\ and\ \citenamefont {San~Juan}}]{Baake:2021jzv}%
  \BibitemOpen
  \bibfield  {author} {\bibinfo {author} {\bibfnamefont {O.}~\bibnamefont {Baake}}, \bibinfo {author} {\bibfnamefont {C.}~\bibnamefont {Charmousis}}, \bibinfo {author} {\bibfnamefont {M.}~\bibnamefont {Hassaine}},\ and\ \bibinfo {author} {\bibfnamefont {M.}~\bibnamefont {San~Juan}},\ }\bibfield  {title} {\bibinfo {title} {{Regular black holes and gravitational particle-like solutions in generic DHOST theories}},\ }\href {https://doi.org/10.1088/1475-7516/2021/06/021} {\bibfield  {journal} {\bibinfo  {journal} {JCAP}\ }\textbf {\bibinfo {volume} {06}},\ \bibinfo {pages} {021}},\ \Eprint {https://arxiv.org/abs/2104.08221} {arXiv:2104.08221 [hep-th]} \BibitemShut {NoStop}%
\bibitem [{\citenamefont {Coll\'eaux}\ \emph {et~al.}(2018)\citenamefont {Coll\'eaux}, \citenamefont {Chinaglia},\ and\ \citenamefont {Zerbini}}]{Colleaux:2017ibe}%
  \BibitemOpen
  \bibfield  {author} {\bibinfo {author} {\bibfnamefont {A.}~\bibnamefont {Coll\'eaux}}, \bibinfo {author} {\bibfnamefont {S.}~\bibnamefont {Chinaglia}},\ and\ \bibinfo {author} {\bibfnamefont {S.}~\bibnamefont {Zerbini}},\ }\bibfield  {title} {\bibinfo {title} {{Nonpolynomial Lagrangian approach to regular black holes}},\ }\href {https://doi.org/10.1142/S0218271818300021} {\bibfield  {journal} {\bibinfo  {journal} {Int. J. Mod. Phys. D}\ }\textbf {\bibinfo {volume} {27}},\ \bibinfo {pages} {1830002} (\bibinfo {year} {2018})},\ \Eprint {https://arxiv.org/abs/1712.03730} {arXiv:1712.03730 [gr-qc]} \BibitemShut {NoStop}%
\bibitem [{\citenamefont {Bueno}\ \emph {et~al.}(2021)\citenamefont {Bueno}, \citenamefont {Cano}, \citenamefont {Moreno},\ and\ \citenamefont {van~der Velde}}]{buenoRegularBlackHoles2021}%
  \BibitemOpen
  \bibfield  {author} {\bibinfo {author} {\bibfnamefont {P.}~\bibnamefont {Bueno}}, \bibinfo {author} {\bibfnamefont {P.~A.}\ \bibnamefont {Cano}}, \bibinfo {author} {\bibfnamefont {J.}~\bibnamefont {Moreno}},\ and\ \bibinfo {author} {\bibfnamefont {G.}~\bibnamefont {van~der Velde}},\ }\bibfield  {title} {\bibinfo {title} {Regular black holes in three dimensions},\ }\href {https://doi.org/10.1103/PhysRevD.104.L021501} {\bibfield  {journal} {\bibinfo  {journal} {Physical Review D}\ }\textbf {\bibinfo {volume} {104}},\ \bibinfo {pages} {L021501} (\bibinfo {year} {2021})},\ \Eprint {https://arxiv.org/abs/2104.10172} {arXiv:2104.10172} \BibitemShut {NoStop}%
\bibitem [{\citenamefont {Bueno}\ \emph {et~al.}(2025{\natexlab{a}})\citenamefont {Bueno}, \citenamefont {Andino}, \citenamefont {Moreno},\ and\ \citenamefont {van~der Velde}}]{buenoRegularChargedBlack2025}%
  \BibitemOpen
  \bibfield  {author} {\bibinfo {author} {\bibfnamefont {P.}~\bibnamefont {Bueno}}, \bibinfo {author} {\bibfnamefont {O.~L.}\ \bibnamefont {Andino}}, \bibinfo {author} {\bibfnamefont {J.}~\bibnamefont {Moreno}},\ and\ \bibinfo {author} {\bibfnamefont {G.}~\bibnamefont {van~der Velde}},\ }\href {https://doi.org/10.48550/arXiv.2503.02930} {\bibinfo {title} {On regular charged black holes in three dimensions}} (\bibinfo {year} {2025}{\natexlab{a}}),\ \Eprint {https://arxiv.org/abs/2503.02930} {arXiv:2503.02930} \BibitemShut {NoStop}%
\bibitem [{\citenamefont {Giacchini}\ and\ \citenamefont {Kol{\'a}{\v{r}}}(2024)}]{Giacchini:2024exc}%
  \BibitemOpen
  \bibfield  {author} {\bibinfo {author} {\bibfnamefont {B.~L.}\ \bibnamefont {Giacchini}}\ and\ \bibinfo {author} {\bibfnamefont {I.}~\bibnamefont {Kol{\'a}{\v{r}}}},\ }\bibfield  {title} {\bibinfo {title} {{Toward regular black holes in sixth-derivative gravity}},\ }\href {https://doi.org/10.1103/PhysRevD.110.104056} {\bibfield  {journal} {\bibinfo  {journal} {Phys. Rev. D}\ }\textbf {\bibinfo {volume} {110}},\ \bibinfo {pages} {104056} (\bibinfo {year} {2024})},\ \Eprint {https://arxiv.org/abs/2406.00997} {arXiv:2406.00997 [gr-qc]} \BibitemShut {NoStop}%
\bibitem [{\citenamefont {Bueno}\ \emph {et~al.}(2025{\natexlab{b}})\citenamefont {Bueno}, \citenamefont {Cano},\ and\ \citenamefont {Hennigar}}]{Bueno:2024dgm}%
  \BibitemOpen
  \bibfield  {author} {\bibinfo {author} {\bibfnamefont {P.}~\bibnamefont {Bueno}}, \bibinfo {author} {\bibfnamefont {P.~A.}\ \bibnamefont {Cano}},\ and\ \bibinfo {author} {\bibfnamefont {R.~A.}\ \bibnamefont {Hennigar}},\ }\bibfield  {title} {\bibinfo {title} {{Regular black holes from pure gravity}},\ }\href {https://doi.org/10.1016/j.physletb.2025.139260} {\bibfield  {journal} {\bibinfo  {journal} {Phys. Lett. B}\ }\textbf {\bibinfo {volume} {861}},\ \bibinfo {pages} {139260} (\bibinfo {year} {2025}{\natexlab{b}})},\ \Eprint {https://arxiv.org/abs/2403.04827} {arXiv:2403.04827 [gr-qc]} \BibitemShut {NoStop}%
\bibitem [{\citenamefont {Bueno}\ \emph {et~al.}(2024{\natexlab{a}})\citenamefont {Bueno}, \citenamefont {Cano}, \citenamefont {Hennigar},\ and\ \citenamefont {Murcia}}]{Bueno:2024zsx}%
  \BibitemOpen
  \bibfield  {author} {\bibinfo {author} {\bibfnamefont {P.}~\bibnamefont {Bueno}}, \bibinfo {author} {\bibfnamefont {P.~A.}\ \bibnamefont {Cano}}, \bibinfo {author} {\bibfnamefont {R.~A.}\ \bibnamefont {Hennigar}},\ and\ \bibinfo {author} {\bibfnamefont {A.~J.}\ \bibnamefont {Murcia}},\ }\href@noop {} {\bibinfo {title} {{Regular black holes from thin-shell collapse}}} (\bibinfo {year} {2024}{\natexlab{a}}),\ \Eprint {https://arxiv.org/abs/2412.02740} {arXiv:2412.02740 [gr-qc]} \BibitemShut {NoStop}%
\bibitem [{\citenamefont {Bueno}\ \emph {et~al.}(2024{\natexlab{b}})\citenamefont {Bueno}, \citenamefont {Cano}, \citenamefont {Hennigar},\ and\ \citenamefont {Murcia}}]{Bueno:2024eig}%
  \BibitemOpen
  \bibfield  {author} {\bibinfo {author} {\bibfnamefont {P.}~\bibnamefont {Bueno}}, \bibinfo {author} {\bibfnamefont {P.~A.}\ \bibnamefont {Cano}}, \bibinfo {author} {\bibfnamefont {R.~A.}\ \bibnamefont {Hennigar}},\ and\ \bibinfo {author} {\bibfnamefont {A.~J.}\ \bibnamefont {Murcia}},\ }\href@noop {} {\bibinfo {title} {{Dynamical Formation of Regular Black Holes}}} (\bibinfo {year} {2024}{\natexlab{b}}),\ \Eprint {https://arxiv.org/abs/2412.02742} {arXiv:2412.02742 [gr-qc]} \BibitemShut {NoStop}%
\bibitem [{\citenamefont {Bueno}\ \emph {et~al.}(2025{\natexlab{c}})\citenamefont {Bueno}, \citenamefont {Cano}, \citenamefont {Hennigar}, \citenamefont {Murcia},\ and\ \citenamefont {Vicente-Cano}}]{Bueno:2025gjg}%
  \BibitemOpen
  \bibfield  {author} {\bibinfo {author} {\bibfnamefont {P.}~\bibnamefont {Bueno}}, \bibinfo {author} {\bibfnamefont {P.~A.}\ \bibnamefont {Cano}}, \bibinfo {author} {\bibfnamefont {R.~A.}\ \bibnamefont {Hennigar}}, \bibinfo {author} {\bibfnamefont {{\'A}.~J.}\ \bibnamefont {Murcia}},\ and\ \bibinfo {author} {\bibfnamefont {A.}~\bibnamefont {Vicente-Cano}},\ }\bibfield  {title} {\bibinfo {title} {{Regular black holes from Oppenheimer-Snyder collapse}},\ }\href {https://doi.org/10.1103/qrbb-mdvm} {\bibfield  {journal} {\bibinfo  {journal} {Phys. Rev. D}\ }\textbf {\bibinfo {volume} {112}},\ \bibinfo {pages} {064039} (\bibinfo {year} {2025}{\natexlab{c}})},\ \Eprint {https://arxiv.org/abs/2505.09680} {arXiv:2505.09680 [gr-qc]} \BibitemShut {NoStop}%
\bibitem [{\citenamefont {Aguayo}\ \emph {et~al.}(2025)\citenamefont {Aguayo}, \citenamefont {Gajardo}, \citenamefont {Grandi}, \citenamefont {Moreno}, \citenamefont {Oliva},\ and\ \citenamefont {Reyes}}]{Aguayo:2025xfi}%
  \BibitemOpen
  \bibfield  {author} {\bibinfo {author} {\bibfnamefont {M.}~\bibnamefont {Aguayo}}, \bibinfo {author} {\bibfnamefont {L.}~\bibnamefont {Gajardo}}, \bibinfo {author} {\bibfnamefont {N.}~\bibnamefont {Grandi}}, \bibinfo {author} {\bibfnamefont {J.}~\bibnamefont {Moreno}}, \bibinfo {author} {\bibfnamefont {J.}~\bibnamefont {Oliva}},\ and\ \bibinfo {author} {\bibfnamefont {M.}~\bibnamefont {Reyes}},\ }\bibfield  {title} {\bibinfo {title} {{Holographic explorations of regular black holes in pure gravity}},\ }\href {https://doi.org/10.1007/JHEP09(2025)030} {\bibfield  {journal} {\bibinfo  {journal} {JHEP}\ }\textbf {\bibinfo {volume} {09}},\ \bibinfo {pages} {030}},\ \Eprint {https://arxiv.org/abs/2505.11736} {arXiv:2505.11736 [hep-th]} \BibitemShut {NoStop}%
\bibitem [{\citenamefont {Hennigar}\ \emph {et~al.}(2025)\citenamefont {Hennigar}, \citenamefont {Kubiz{\v{n}}{\'a}k}, \citenamefont {Murk},\ and\ \citenamefont {Soranidis}}]{Hennigar:2025ftm}%
  \BibitemOpen
  \bibfield  {author} {\bibinfo {author} {\bibfnamefont {R.~A.}\ \bibnamefont {Hennigar}}, \bibinfo {author} {\bibfnamefont {D.}~\bibnamefont {Kubiz{\v{n}}{\'a}k}}, \bibinfo {author} {\bibfnamefont {S.}~\bibnamefont {Murk}},\ and\ \bibinfo {author} {\bibfnamefont {I.}~\bibnamefont {Soranidis}},\ }\href@noop {} {\bibinfo {title} {{Thermodynamics of Regular Black Holes in Anti-de Sitter Space}}} (\bibinfo {year} {2025}),\ \Eprint {https://arxiv.org/abs/2505.11623} {arXiv:2505.11623 [gr-qc]} \BibitemShut {NoStop}%
\bibitem [{\citenamefont {Konoplya}\ and\ \citenamefont {Zhidenko}(2024{\natexlab{a}})}]{Konoplya:2024hfg}%
  \BibitemOpen
  \bibfield  {author} {\bibinfo {author} {\bibfnamefont {R.~A.}\ \bibnamefont {Konoplya}}\ and\ \bibinfo {author} {\bibfnamefont {A.}~\bibnamefont {Zhidenko}},\ }\bibfield  {title} {\bibinfo {title} {{Infinite tower of higher-curvature corrections: Quasinormal modes and late-time behavior of D-dimensional regular black holes}},\ }\href {https://doi.org/10.1103/PhysRevD.109.104005} {\bibfield  {journal} {\bibinfo  {journal} {Phys. Rev. D}\ }\textbf {\bibinfo {volume} {109}},\ \bibinfo {pages} {104005} (\bibinfo {year} {2024}{\natexlab{a}})},\ \Eprint {https://arxiv.org/abs/2403.07848} {arXiv:2403.07848 [gr-qc]} \BibitemShut {NoStop}%
\bibitem [{\citenamefont {Di~Filippo}\ \emph {et~al.}(2025)\citenamefont {Di~Filippo}, \citenamefont {Kol{\'a}{\v{r}}},\ and\ \citenamefont {Kubiznak}}]{DiFilippo:2024mwm}%
  \BibitemOpen
  \bibfield  {author} {\bibinfo {author} {\bibfnamefont {F.}~\bibnamefont {Di~Filippo}}, \bibinfo {author} {\bibfnamefont {I.}~\bibnamefont {Kol{\'a}{\v{r}}}},\ and\ \bibinfo {author} {\bibfnamefont {D.}~\bibnamefont {Kubiznak}},\ }\bibfield  {title} {\bibinfo {title} {{Inner-extremal regular black holes from pure gravity}},\ }\href {https://doi.org/10.1103/PhysRevD.111.L041505} {\bibfield  {journal} {\bibinfo  {journal} {Phys. Rev. D}\ }\textbf {\bibinfo {volume} {111}},\ \bibinfo {pages} {L041505} (\bibinfo {year} {2025})},\ \Eprint {https://arxiv.org/abs/2404.07058} {arXiv:2404.07058 [gr-qc]} \BibitemShut {NoStop}%
\bibitem [{\citenamefont {Konoplya}\ and\ \citenamefont {Zhidenko}(2024{\natexlab{b}})}]{Konoplya:2024kih}%
  \BibitemOpen
  \bibfield  {author} {\bibinfo {author} {\bibfnamefont {R.~A.}\ \bibnamefont {Konoplya}}\ and\ \bibinfo {author} {\bibfnamefont {A.}~\bibnamefont {Zhidenko}},\ }\bibfield  {title} {\bibinfo {title} {{Dymnikova black hole from an infinite tower of higher-curvature corrections}},\ }\href {https://doi.org/10.1016/j.physletb.2024.138945} {\bibfield  {journal} {\bibinfo  {journal} {Phys. Lett. B}\ }\textbf {\bibinfo {volume} {856}},\ \bibinfo {pages} {138945} (\bibinfo {year} {2024}{\natexlab{b}})},\ \Eprint {https://arxiv.org/abs/2404.09063} {arXiv:2404.09063 [gr-qc]} \BibitemShut {NoStop}%
\bibitem [{\citenamefont {Cisterna}\ \emph {et~al.}(2024)\citenamefont {Cisterna}, \citenamefont {Grandi},\ and\ \citenamefont {Oliva}}]{Cisterna:2024ksz}%
  \BibitemOpen
  \bibfield  {author} {\bibinfo {author} {\bibfnamefont {A.}~\bibnamefont {Cisterna}}, \bibinfo {author} {\bibfnamefont {N.}~\bibnamefont {Grandi}},\ and\ \bibinfo {author} {\bibfnamefont {J.}~\bibnamefont {Oliva}},\ }\bibfield  {title} {\bibinfo {title} {{de Sitter geometric inflation from dynamical singularities}},\ }\href {https://doi.org/10.1103/PhysRevD.110.084043} {\bibfield  {journal} {\bibinfo  {journal} {Phys. Rev. D}\ }\textbf {\bibinfo {volume} {110}},\ \bibinfo {pages} {084043} (\bibinfo {year} {2024})},\ \Eprint {https://arxiv.org/abs/2406.10037} {arXiv:2406.10037 [hep-th]} \BibitemShut {NoStop}%
\bibitem [{\citenamefont {Frolov}\ \emph {et~al.}(2025)\citenamefont {Frolov}, \citenamefont {Koek}, \citenamefont {Soto},\ and\ \citenamefont {Zelnikov}}]{Frolov:2024hhe}%
  \BibitemOpen
  \bibfield  {author} {\bibinfo {author} {\bibfnamefont {V.~P.}\ \bibnamefont {Frolov}}, \bibinfo {author} {\bibfnamefont {A.}~\bibnamefont {Koek}}, \bibinfo {author} {\bibfnamefont {J.~P.}\ \bibnamefont {Soto}},\ and\ \bibinfo {author} {\bibfnamefont {A.}~\bibnamefont {Zelnikov}},\ }\bibfield  {title} {\bibinfo {title} {{Regular black holes inspired by quasitopological gravity}},\ }\href {https://doi.org/10.1103/PhysRevD.111.044034} {\bibfield  {journal} {\bibinfo  {journal} {Phys. Rev. D}\ }\textbf {\bibinfo {volume} {111}},\ \bibinfo {pages} {044034} (\bibinfo {year} {2025})},\ \Eprint {https://arxiv.org/abs/2411.16050} {arXiv:2411.16050 [gr-qc]} \BibitemShut {NoStop}%
\bibitem [{\citenamefont {Srivastava}\ \emph {et~al.}(2025)\citenamefont {Srivastava}, \citenamefont {Upadhyay}, \citenamefont {Verma}, \citenamefont {Singh}, \citenamefont {Myrzakulov},\ and\ \citenamefont {Myrzakulov}}]{Srivastava:2025rxe}%
  \BibitemOpen
  \bibfield  {author} {\bibinfo {author} {\bibfnamefont {V.~K.}\ \bibnamefont {Srivastava}}, \bibinfo {author} {\bibfnamefont {S.}~\bibnamefont {Upadhyay}}, \bibinfo {author} {\bibfnamefont {A.~K.}\ \bibnamefont {Verma}}, \bibinfo {author} {\bibfnamefont {D.~V.}\ \bibnamefont {Singh}}, \bibinfo {author} {\bibfnamefont {Y.}~\bibnamefont {Myrzakulov}},\ and\ \bibinfo {author} {\bibfnamefont {K.}~\bibnamefont {Myrzakulov}},\ }\bibfield  {title} {\bibinfo {title} {{Exploring non-perturbative effects on quasi-topological black hole thermodynamics}},\ }\href {https://doi.org/10.1016/j.dark.2025.101915} {\bibfield  {journal} {\bibinfo  {journal} {Phys. Dark Univ.}\ }\textbf {\bibinfo {volume} {48}},\ \bibinfo {pages} {101915} (\bibinfo {year} {2025})},\ \Eprint {https://arxiv.org/abs/2504.15318} {arXiv:2504.15318 [gr-qc]} \BibitemShut {NoStop}%
\bibitem [{\citenamefont {Konoplya}\ and\ \citenamefont {Zhidenko}(2025{\natexlab{a}})}]{Konoplya:2025uta}%
  \BibitemOpen
  \bibfield  {author} {\bibinfo {author} {\bibfnamefont {R.~A.}\ \bibnamefont {Konoplya}}\ and\ \bibinfo {author} {\bibfnamefont {A.}~\bibnamefont {Zhidenko}},\ }\bibfield  {title} {\bibinfo {title} {{Convergence of Higher-Curvature Expansions Near the Horizon: Hawking Radiation from Regular Black Holes}},\ }\href {https://doi.org/10.53941/ijgtp.2025.100005} {\bibfield  {journal} {\bibinfo  {journal} {Theor. Phys.}\ }\textbf {\bibinfo {volume} {1}},\ \bibinfo {pages} {5} (\bibinfo {year} {2025}{\natexlab{a}})},\ \Eprint {https://arxiv.org/abs/2507.22660} {arXiv:2507.22660 [gr-qc]} \BibitemShut {NoStop}%
\bibitem [{\citenamefont {Arbelaez}(2025)}]{Arbelaez:2025gwj}%
  \BibitemOpen
  \bibfield  {author} {\bibinfo {author} {\bibfnamefont {J.~P.}\ \bibnamefont {Arbelaez}},\ }\href@noop {} {\bibinfo {title} {{Quasinormal spectra of higher dimensional regular black holes in theories with infinite curvature corrections}}} (\bibinfo {year} {2025}),\ \Eprint {https://arxiv.org/abs/2509.25141} {arXiv:2509.25141 [gr-qc]} \BibitemShut {NoStop}%
\bibitem [{\citenamefont {Fernandes}(2025{\natexlab{a}})}]{Fernandes:2025fnz}%
  \BibitemOpen
  \bibfield  {author} {\bibinfo {author} {\bibfnamefont {P.~G.~S.}\ \bibnamefont {Fernandes}},\ }\bibfield  {title} {\bibinfo {title} {{Singularity resolution and inflation from an infinite tower of regularized curvature corrections}},\ }\href {https://doi.org/10.1103/763p-htct} {\bibfield  {journal} {\bibinfo  {journal} {Phys. Rev. D}\ }\textbf {\bibinfo {volume} {112}},\ \bibinfo {pages} {084028} (\bibinfo {year} {2025}{\natexlab{a}})},\ \Eprint {https://arxiv.org/abs/2504.07692} {arXiv:2504.07692 [gr-qc]} \BibitemShut {NoStop}%
\bibitem [{\citenamefont {Fernandes}(2025{\natexlab{b}})}]{Fernandes:2025eoc}%
  \BibitemOpen
  \bibfield  {author} {\bibinfo {author} {\bibfnamefont {P.~G.~S.}\ \bibnamefont {Fernandes}},\ }\bibfield  {title} {\bibinfo {title} {{Regular BTZ black holes from an infinite tower of corrections}},\ }\href {https://doi.org/10.1016/j.physletb.2025.139772} {\bibfield  {journal} {\bibinfo  {journal} {Phys. Lett. B}\ }\textbf {\bibinfo {volume} {868}},\ \bibinfo {pages} {139772} (\bibinfo {year} {2025}{\natexlab{b}})},\ \Eprint {https://arxiv.org/abs/2504.08565} {arXiv:2504.08565 [gr-qc]} \BibitemShut {NoStop}%
\bibitem [{\citenamefont {Bueno}\ \emph {et~al.}(2025{\natexlab{d}})\citenamefont {Bueno}, \citenamefont {Cano}, \citenamefont {Hennigar},\ and\ \citenamefont {Murcia}}]{Bueno:2025zaj}%
  \BibitemOpen
  \bibfield  {author} {\bibinfo {author} {\bibfnamefont {P.}~\bibnamefont {Bueno}}, \bibinfo {author} {\bibfnamefont {P.~A.}\ \bibnamefont {Cano}}, \bibinfo {author} {\bibfnamefont {R.~A.}\ \bibnamefont {Hennigar}},\ and\ \bibinfo {author} {\bibfnamefont {{\'A}.~J.}\ \bibnamefont {Murcia}},\ }\href@noop {} {\bibinfo {title} {{Regular black hole formation in four-dimensional non-polynomial gravities}}} (\bibinfo {year} {2025}{\natexlab{d}}),\ \Eprint {https://arxiv.org/abs/2509.19016} {arXiv:2509.19016 [gr-qc]} \BibitemShut {NoStop}%
\bibitem [{\citenamefont {Fernandes}\ \emph {et~al.}(2022)\citenamefont {Fernandes}, \citenamefont {Carrilho}, \citenamefont {Clifton},\ and\ \citenamefont {Mulryne}}]{Fernandes:2022zrq}%
  \BibitemOpen
  \bibfield  {author} {\bibinfo {author} {\bibfnamefont {P.~G.~S.}\ \bibnamefont {Fernandes}}, \bibinfo {author} {\bibfnamefont {P.}~\bibnamefont {Carrilho}}, \bibinfo {author} {\bibfnamefont {T.}~\bibnamefont {Clifton}},\ and\ \bibinfo {author} {\bibfnamefont {D.~J.}\ \bibnamefont {Mulryne}},\ }\bibfield  {title} {\bibinfo {title} {{The 4D Einstein-Gauss-Bonnet theory of gravity: a review}},\ }\href {https://doi.org/10.1088/1361-6382/ac500a} {\bibfield  {journal} {\bibinfo  {journal} {Class. Quant. Grav.}\ }\textbf {\bibinfo {volume} {39}},\ \bibinfo {pages} {063001} (\bibinfo {year} {2022})},\ \Eprint {https://arxiv.org/abs/2202.13908} {arXiv:2202.13908 [gr-qc]} \BibitemShut {NoStop}%
\bibitem [{\citenamefont {Glavan}\ and\ \citenamefont {Lin}(2020)}]{Glavan:2019inb}%
  \BibitemOpen
  \bibfield  {author} {\bibinfo {author} {\bibfnamefont {D.}~\bibnamefont {Glavan}}\ and\ \bibinfo {author} {\bibfnamefont {C.}~\bibnamefont {Lin}},\ }\bibfield  {title} {\bibinfo {title} {{Einstein-Gauss-Bonnet Gravity in Four-Dimensional Spacetime}},\ }\href {https://doi.org/10.1103/PhysRevLett.124.081301} {\bibfield  {journal} {\bibinfo  {journal} {Phys. Rev. Lett.}\ }\textbf {\bibinfo {volume} {124}},\ \bibinfo {pages} {081301} (\bibinfo {year} {2020})},\ \Eprint {https://arxiv.org/abs/1905.03601} {arXiv:1905.03601 [gr-qc]} \BibitemShut {NoStop}%
\bibitem [{\citenamefont {Fernandes}\ \emph {et~al.}(2020)\citenamefont {Fernandes}, \citenamefont {Carrilho}, \citenamefont {Clifton},\ and\ \citenamefont {Mulryne}}]{Fernandes:2020nbq}%
  \BibitemOpen
  \bibfield  {author} {\bibinfo {author} {\bibfnamefont {P.~G.~S.}\ \bibnamefont {Fernandes}}, \bibinfo {author} {\bibfnamefont {P.}~\bibnamefont {Carrilho}}, \bibinfo {author} {\bibfnamefont {T.}~\bibnamefont {Clifton}},\ and\ \bibinfo {author} {\bibfnamefont {D.~J.}\ \bibnamefont {Mulryne}},\ }\bibfield  {title} {\bibinfo {title} {{Derivation of Regularized Field Equations for the Einstein-Gauss-Bonnet Theory in Four Dimensions}},\ }\href {https://doi.org/10.1103/PhysRevD.102.024025} {\bibfield  {journal} {\bibinfo  {journal} {Phys. Rev. D}\ }\textbf {\bibinfo {volume} {102}},\ \bibinfo {pages} {024025} (\bibinfo {year} {2020})},\ \Eprint {https://arxiv.org/abs/2004.08362} {arXiv:2004.08362 [gr-qc]} \BibitemShut {NoStop}%
\bibitem [{\citenamefont {Hennigar}\ \emph {et~al.}(2020)\citenamefont {Hennigar}, \citenamefont {Kubiz\v{n}\'ak}, \citenamefont {Mann},\ and\ \citenamefont {Pollack}}]{Hennigar:2020lsl}%
  \BibitemOpen
  \bibfield  {author} {\bibinfo {author} {\bibfnamefont {R.~A.}\ \bibnamefont {Hennigar}}, \bibinfo {author} {\bibfnamefont {D.}~\bibnamefont {Kubiz\v{n}\'ak}}, \bibinfo {author} {\bibfnamefont {R.~B.}\ \bibnamefont {Mann}},\ and\ \bibinfo {author} {\bibfnamefont {C.}~\bibnamefont {Pollack}},\ }\bibfield  {title} {\bibinfo {title} {{On taking the D \textrightarrow{} 4 limit of Gauss-Bonnet gravity: theory and solutions}},\ }\href {https://doi.org/10.1007/JHEP07(2020)027} {\bibfield  {journal} {\bibinfo  {journal} {JHEP}\ }\textbf {\bibinfo {volume} {07}},\ \bibinfo {pages} {027}},\ \Eprint {https://arxiv.org/abs/2004.09472} {arXiv:2004.09472 [gr-qc]} \BibitemShut {NoStop}%
\bibitem [{\citenamefont {Lu}\ and\ \citenamefont {Pang}(2020)}]{Lu:2020iav}%
  \BibitemOpen
  \bibfield  {author} {\bibinfo {author} {\bibfnamefont {H.}~\bibnamefont {Lu}}\ and\ \bibinfo {author} {\bibfnamefont {Y.}~\bibnamefont {Pang}},\ }\bibfield  {title} {\bibinfo {title} {{Horndeski gravity as $D \rightarrow 4$ limit of Gauss-Bonnet}},\ }\href {https://doi.org/10.1016/j.physletb.2020.135717} {\bibfield  {journal} {\bibinfo  {journal} {Phys. Lett. B}\ }\textbf {\bibinfo {volume} {809}},\ \bibinfo {pages} {135717} (\bibinfo {year} {2020})},\ \Eprint {https://arxiv.org/abs/2003.11552} {arXiv:2003.11552 [gr-qc]} \BibitemShut {NoStop}%
\bibitem [{\citenamefont {Kobayashi}(2020)}]{Kobayashi:2020wqy}%
  \BibitemOpen
  \bibfield  {author} {\bibinfo {author} {\bibfnamefont {T.}~\bibnamefont {Kobayashi}},\ }\bibfield  {title} {\bibinfo {title} {{Effective scalar-tensor description of regularized Lovelock gravity in four dimensions}},\ }\href {https://doi.org/10.1088/1475-7516/2020/07/013} {\bibfield  {journal} {\bibinfo  {journal} {JCAP}\ }\textbf {\bibinfo {volume} {07}},\ \bibinfo {pages} {013}},\ \Eprint {https://arxiv.org/abs/2003.12771} {arXiv:2003.12771 [gr-qc]} \BibitemShut {NoStop}%
\bibitem [{\citenamefont {Fernandes}(2021)}]{Fernandes:2021dsb}%
  \BibitemOpen
  \bibfield  {author} {\bibinfo {author} {\bibfnamefont {P.~G.~S.}\ \bibnamefont {Fernandes}},\ }\bibfield  {title} {\bibinfo {title} {{Gravity with a generalized conformal scalar field: theory and solutions}},\ }\href {https://doi.org/10.1103/PhysRevD.103.104065} {\bibfield  {journal} {\bibinfo  {journal} {Phys. Rev. D}\ }\textbf {\bibinfo {volume} {103}},\ \bibinfo {pages} {104065} (\bibinfo {year} {2021})},\ \Eprint {https://arxiv.org/abs/2105.04687} {arXiv:2105.04687 [gr-qc]} \BibitemShut {NoStop}%
\bibitem [{\citenamefont {Bonifacio}\ \emph {et~al.}(2020)\citenamefont {Bonifacio}, \citenamefont {Hinterbichler},\ and\ \citenamefont {Johnson}}]{Bonifacio:2020vbk}%
  \BibitemOpen
  \bibfield  {author} {\bibinfo {author} {\bibfnamefont {J.}~\bibnamefont {Bonifacio}}, \bibinfo {author} {\bibfnamefont {K.}~\bibnamefont {Hinterbichler}},\ and\ \bibinfo {author} {\bibfnamefont {L.~A.}\ \bibnamefont {Johnson}},\ }\bibfield  {title} {\bibinfo {title} {{Amplitudes and 4D Gauss-Bonnet Theory}},\ }\href {https://doi.org/10.1103/PhysRevD.102.024029} {\bibfield  {journal} {\bibinfo  {journal} {Phys. Rev. D}\ }\textbf {\bibinfo {volume} {102}},\ \bibinfo {pages} {024029} (\bibinfo {year} {2020})},\ \Eprint {https://arxiv.org/abs/2004.10716} {arXiv:2004.10716 [hep-th]} \BibitemShut {NoStop}%
\bibitem [{\citenamefont {Tsujikawa}(2025)}]{Tsujikawa:2025eac}%
  \BibitemOpen
  \bibfield  {author} {\bibinfo {author} {\bibfnamefont {S.}~\bibnamefont {Tsujikawa}},\ }\bibfield  {title} {\bibinfo {title} {{Strong coupling and instabilities in singularity-free inflation from an infinite sum of curvature corrections}},\ }\href {https://doi.org/10.1103/9x4j-hxwx} {\bibfield  {journal} {\bibinfo  {journal} {Phys. Rev. D}\ }\textbf {\bibinfo {volume} {112}},\ \bibinfo {pages} {024062} (\bibinfo {year} {2025})},\ \Eprint {https://arxiv.org/abs/2505.20586} {arXiv:2505.20586 [gr-qc]} \BibitemShut {NoStop}%
\bibitem [{\citenamefont {Cisterna}\ \emph {et~al.}(2025)\citenamefont {Cisterna}, \citenamefont {Hassaine},\ and\ \citenamefont {Hernandez-Vera}}]{Cisterna:2025vxk}%
  \BibitemOpen
  \bibfield  {author} {\bibinfo {author} {\bibfnamefont {A.}~\bibnamefont {Cisterna}}, \bibinfo {author} {\bibfnamefont {M.}~\bibnamefont {Hassaine}},\ and\ \bibinfo {author} {\bibfnamefont {U.}~\bibnamefont {Hernandez-Vera}},\ }\bibfield  {title} {\bibinfo {title} {{Thermodynamics of four-dimensional regular black holes with an infinite tower of regularized curvature corrections}},\ }\href {https://doi.org/10.1103/6f3b-8794} {\bibfield  {journal} {\bibinfo  {journal} {Phys. Rev. D}\ }\textbf {\bibinfo {volume} {112}},\ \bibinfo {pages} {064036} (\bibinfo {year} {2025})},\ \Eprint {https://arxiv.org/abs/2505.23467} {arXiv:2505.23467 [gr-qc]} \BibitemShut {NoStop}%
\bibitem [{\citenamefont {De~Felice}\ and\ \citenamefont {Tsujikawa}(2025)}]{DeFelice:2025fzv}%
  \BibitemOpen
  \bibfield  {author} {\bibinfo {author} {\bibfnamefont {A.}~\bibnamefont {De~Felice}}\ and\ \bibinfo {author} {\bibfnamefont {S.}~\bibnamefont {Tsujikawa}},\ }\bibfield  {title} {\bibinfo {title} {{Instability of regular planar black holes in four dimensions arising from an infinite sum of curvature corrections}},\ }\href {https://doi.org/10.1103/cbyk-b46n} {\bibfield  {journal} {\bibinfo  {journal} {Phys. Rev. D}\ }\textbf {\bibinfo {volume} {112}},\ \bibinfo {pages} {064023} (\bibinfo {year} {2025})},\ \Eprint {https://arxiv.org/abs/2507.11803} {arXiv:2507.11803 [gr-qc]} \BibitemShut {NoStop}%
\bibitem [{\citenamefont {Ling}\ and\ \citenamefont {Yu}(2025)}]{Ling:2025ncw}%
  \BibitemOpen
  \bibfield  {author} {\bibinfo {author} {\bibfnamefont {Y.}~\bibnamefont {Ling}}\ and\ \bibinfo {author} {\bibfnamefont {Z.}~\bibnamefont {Yu}},\ }\href@noop {} {\bibinfo {title} {{Big bounce and black bounce in quasi-topological gravity}}} (\bibinfo {year} {2025}),\ \Eprint {https://arxiv.org/abs/2509.00137} {arXiv:2509.00137 [gr-qc]} \BibitemShut {NoStop}%
\bibitem [{\citenamefont {Charmousis}\ \emph {et~al.}(2025)\citenamefont {Charmousis}, \citenamefont {Fernandes},\ and\ \citenamefont {Hassaine}}]{Charmousis:2025jpx}%
  \BibitemOpen
  \bibfield  {author} {\bibinfo {author} {\bibfnamefont {C.}~\bibnamefont {Charmousis}}, \bibinfo {author} {\bibfnamefont {P.~G.~S.}\ \bibnamefont {Fernandes}},\ and\ \bibinfo {author} {\bibfnamefont {M.}~\bibnamefont {Hassaine}},\ }\bibfield  {title} {\bibinfo {title} {{Proca theory of four-dimensional regularized Gauss-Bonnet gravity and black holes with primary hair}},\ }\href {https://doi.org/10.1103/9f2w-3kly} {\bibfield  {journal} {\bibinfo  {journal} {Phys. Rev. D}\ }\textbf {\bibinfo {volume} {111}},\ \bibinfo {pages} {124008} (\bibinfo {year} {2025})},\ \Eprint {https://arxiv.org/abs/2504.13084} {arXiv:2504.13084 [gr-qc]} \BibitemShut {NoStop}%
\bibitem [{\citenamefont {Heisenberg}(2014)}]{Heisenberg:2014rta}%
  \BibitemOpen
  \bibfield  {author} {\bibinfo {author} {\bibfnamefont {L.}~\bibnamefont {Heisenberg}},\ }\bibfield  {title} {\bibinfo {title} {{Generalization of the Proca Action}},\ }\href {https://doi.org/10.1088/1475-7516/2014/05/015} {\bibfield  {journal} {\bibinfo  {journal} {JCAP}\ }\textbf {\bibinfo {volume} {05}},\ \bibinfo {pages} {015}},\ \Eprint {https://arxiv.org/abs/1402.7026} {arXiv:1402.7026 [hep-th]} \BibitemShut {NoStop}%
\bibitem [{\citenamefont {Eichhorn}\ and\ \citenamefont {Fernandes}(2025)}]{Eichhorn:2025pgy}%
  \BibitemOpen
  \bibfield  {author} {\bibinfo {author} {\bibfnamefont {A.}~\bibnamefont {Eichhorn}}\ and\ \bibinfo {author} {\bibfnamefont {P.~G.~S.}\ \bibnamefont {Fernandes}},\ }\href@noop {} {\bibinfo {title} {{Regular black holes without mass-inflation instability and gravastars from modified gravity}}} (\bibinfo {year} {2025}),\ \Eprint {https://arxiv.org/abs/2508.00686} {arXiv:2508.00686 [gr-qc]} \BibitemShut {NoStop}%
\bibitem [{\citenamefont {L{\"u}tf{\"u}o{\u{g}}lu}(2025{\natexlab{a}})}]{Lutfuoglu:2025ldc}%
  \BibitemOpen
  \bibfield  {author} {\bibinfo {author} {\bibfnamefont {B.~C.}\ \bibnamefont {L{\"u}tf{\"u}o{\u{g}}lu}},\ }\bibfield  {title} {\bibinfo {title} {{Black Holes in Proca-Gauss-Bonnet Gravity with Primary Hair: Particle Motion, Shadows, and Grey-Body Factors}},\ }\href {https://doi.org/10.53941/ijgtp.2025.100004} {\bibfield  {journal} {\bibinfo  {journal} {Theor. Phys.}\ }\textbf {\bibinfo {volume} {1}},\ \bibinfo {pages} {4} (\bibinfo {year} {2025}{\natexlab{a}})},\ \Eprint {https://arxiv.org/abs/2507.09246} {arXiv:2507.09246 [gr-qc]} \BibitemShut {NoStop}%
\bibitem [{\citenamefont {Alkac}\ \emph {et~al.}(2025{\natexlab{a}})\citenamefont {Alkac}, \citenamefont {Mesta},\ and\ \citenamefont {Unal}}]{Alkac:2025zzi}%
  \BibitemOpen
  \bibfield  {author} {\bibinfo {author} {\bibfnamefont {G.}~\bibnamefont {Alkac}}, \bibinfo {author} {\bibfnamefont {M.}~\bibnamefont {Mesta}},\ and\ \bibinfo {author} {\bibfnamefont {G.}~\bibnamefont {Unal}},\ }\bibfield  {title} {\bibinfo {title} {{AdS3 black holes with primary Proca hair from a regularized Gauss-Bonnet coupling}},\ }\href {https://doi.org/10.1103/1ws1-x8z3} {\bibfield  {journal} {\bibinfo  {journal} {Phys. Rev. D}\ }\textbf {\bibinfo {volume} {112}},\ \bibinfo {pages} {084055} (\bibinfo {year} {2025}{\natexlab{a}})},\ \Eprint {https://arxiv.org/abs/2508.03386} {arXiv:2508.03386 [hep-th]} \BibitemShut {NoStop}%
\bibitem [{\citenamefont {Liu}\ \emph {et~al.}(2025)\citenamefont {Liu}, \citenamefont {Yang}, \citenamefont {Zhu},\ and\ \citenamefont {Liu}}]{Liu:2025dqg}%
  \BibitemOpen
  \bibfield  {author} {\bibinfo {author} {\bibfnamefont {J.-Z.}\ \bibnamefont {Liu}}, \bibinfo {author} {\bibfnamefont {S.-J.}\ \bibnamefont {Yang}}, \bibinfo {author} {\bibfnamefont {C.-C.}\ \bibnamefont {Zhu}},\ and\ \bibinfo {author} {\bibfnamefont {Y.-X.}\ \bibnamefont {Liu}},\ }\href@noop {} {\bibinfo {title} {{D-dimensional black holes in extended Gauss-Bonnet gravity}}} (\bibinfo {year} {2025}),\ \Eprint {https://arxiv.org/abs/2508.04292} {arXiv:2508.04292 [gr-qc]} \BibitemShut {NoStop}%
\bibitem [{\citenamefont {Konoplya}\ and\ \citenamefont {Zhidenko}(2025{\natexlab{b}})}]{Konoplya:2025uiq}%
  \BibitemOpen
  \bibfield  {author} {\bibinfo {author} {\bibfnamefont {R.~A.}\ \bibnamefont {Konoplya}}\ and\ \bibinfo {author} {\bibfnamefont {A.}~\bibnamefont {Zhidenko}},\ }\href@noop {} {\bibinfo {title} {{Primary hairs may create echoes}}} (\bibinfo {year} {2025}{\natexlab{b}}),\ \Eprint {https://arxiv.org/abs/2508.13069} {arXiv:2508.13069 [gr-qc]} \BibitemShut {NoStop}%
\bibitem [{\citenamefont {Alkac}\ \emph {et~al.}(2025{\natexlab{b}})\citenamefont {Alkac}, \citenamefont {Mesta},\ and\ \citenamefont {Unal}}]{Alkac:2025jhx}%
  \BibitemOpen
  \bibfield  {author} {\bibinfo {author} {\bibfnamefont {G.}~\bibnamefont {Alkac}}, \bibinfo {author} {\bibfnamefont {M.}~\bibnamefont {Mesta}},\ and\ \bibinfo {author} {\bibfnamefont {G.}~\bibnamefont {Unal}},\ }\href@noop {} {\bibinfo {title} {{Regular AdS$_3$ black holes from regularized Gauss-Bonnet coupling}}} (\bibinfo {year} {2025}{\natexlab{b}}),\ \Eprint {https://arxiv.org/abs/2508.14010} {arXiv:2508.14010 [hep-th]} \BibitemShut {NoStop}%
\bibitem [{\citenamefont {L{\"u}tf{\"u}o{\u{g}}lu}(2025{\natexlab{b}})}]{Lutfuoglu:2025qkt}%
  \BibitemOpen
  \bibfield  {author} {\bibinfo {author} {\bibfnamefont {B.~C.}\ \bibnamefont {L{\"u}tf{\"u}o{\u{g}}lu}},\ }\bibfield  {title} {\bibinfo {title} {{Long-lived quasinormal modes and echoes in the Einstein{\textendash}Gauss{\textendash}Bonnet{\textendash}Proca theory}},\ }\href {https://doi.org/10.1140/epjc/s10052-025-14839-x} {\bibfield  {journal} {\bibinfo  {journal} {Eur. Phys. J. C}\ }\textbf {\bibinfo {volume} {85}},\ \bibinfo {pages} {1076} (\bibinfo {year} {2025}{\natexlab{b}})},\ \Eprint {https://arxiv.org/abs/2508.19194} {arXiv:2508.19194 [gr-qc]} \BibitemShut {NoStop}%
\bibitem [{\citenamefont {Konoplya}\ \emph {et~al.}(2025)\citenamefont {Konoplya}, \citenamefont {Ovchinnikov},\ and\ \citenamefont {Schee}}]{Konoplya:2025bte}%
  \BibitemOpen
  \bibfield  {author} {\bibinfo {author} {\bibfnamefont {R.~A.}\ \bibnamefont {Konoplya}}, \bibinfo {author} {\bibfnamefont {D.}~\bibnamefont {Ovchinnikov}},\ and\ \bibinfo {author} {\bibfnamefont {J.}~\bibnamefont {Schee}},\ }\href@noop {} {\bibinfo {title} {{Primary Proca Hair and the Double-Peak Optics of Black Holes}}} (\bibinfo {year} {2025}),\ \Eprint {https://arxiv.org/abs/2510.05947} {arXiv:2510.05947 [gr-qc]} \BibitemShut {NoStop}%
\bibitem [{\citenamefont {Poisson}\ and\ \citenamefont {Israel}(1989)}]{PhysRevLett.63.1663}%
  \BibitemOpen
  \bibfield  {author} {\bibinfo {author} {\bibfnamefont {E.}~\bibnamefont {Poisson}}\ and\ \bibinfo {author} {\bibfnamefont {W.}~\bibnamefont {Israel}},\ }\bibfield  {title} {\bibinfo {title} {Inner-horizon instability and mass inflation in black holes},\ }\href {https://doi.org/10.1103/PhysRevLett.63.1663} {\bibfield  {journal} {\bibinfo  {journal} {Phys. Rev. Lett.}\ }\textbf {\bibinfo {volume} {63}},\ \bibinfo {pages} {1663} (\bibinfo {year} {1989})}\BibitemShut {NoStop}%
\bibitem [{\citenamefont {Ori}(1991)}]{PhysRevLett.67.789}%
  \BibitemOpen
  \bibfield  {author} {\bibinfo {author} {\bibfnamefont {A.}~\bibnamefont {Ori}},\ }\bibfield  {title} {\bibinfo {title} {Inner structure of a charged black hole: An exact mass-inflation solution},\ }\href {https://doi.org/10.1103/PhysRevLett.67.789} {\bibfield  {journal} {\bibinfo  {journal} {Phys. Rev. Lett.}\ }\textbf {\bibinfo {volume} {67}},\ \bibinfo {pages} {789} (\bibinfo {year} {1991})}\BibitemShut {NoStop}%
\bibitem [{\citenamefont {Hamilton}\ and\ \citenamefont {Avelino}(2010)}]{Hamilton:2008zz}%
  \BibitemOpen
  \bibfield  {author} {\bibinfo {author} {\bibfnamefont {A.~J.~S.}\ \bibnamefont {Hamilton}}\ and\ \bibinfo {author} {\bibfnamefont {P.~P.}\ \bibnamefont {Avelino}},\ }\bibfield  {title} {\bibinfo {title} {{The Physics of the relativistic counter-streaming instability that drives mass inflation inside black holes}},\ }\href {https://doi.org/10.1016/j.physrep.2010.06.002} {\bibfield  {journal} {\bibinfo  {journal} {Phys. Rept.}\ }\textbf {\bibinfo {volume} {495}},\ \bibinfo {pages} {1} (\bibinfo {year} {2010})},\ \Eprint {https://arxiv.org/abs/0811.1926} {arXiv:0811.1926 [gr-qc]} \BibitemShut {NoStop}%
\bibitem [{\citenamefont {Brown}\ \emph {et~al.}(2011)\citenamefont {Brown}, \citenamefont {Mann},\ and\ \citenamefont {Modesto}}]{Brown:2011tv}%
  \BibitemOpen
  \bibfield  {author} {\bibinfo {author} {\bibfnamefont {E.~G.}\ \bibnamefont {Brown}}, \bibinfo {author} {\bibfnamefont {R.~B.}\ \bibnamefont {Mann}},\ and\ \bibinfo {author} {\bibfnamefont {L.}~\bibnamefont {Modesto}},\ }\bibfield  {title} {\bibinfo {title} {{Mass Inflation in the Loop Black Hole}},\ }\href {https://doi.org/10.1103/PhysRevD.84.104041} {\bibfield  {journal} {\bibinfo  {journal} {Phys. Rev. D}\ }\textbf {\bibinfo {volume} {84}},\ \bibinfo {pages} {104041} (\bibinfo {year} {2011})},\ \Eprint {https://arxiv.org/abs/1104.3126} {arXiv:1104.3126 [gr-qc]} \BibitemShut {NoStop}%
\bibitem [{\citenamefont {Bertipagani}\ \emph {et~al.}(2021)\citenamefont {Bertipagani}, \citenamefont {Rinaldi}, \citenamefont {Sebastiani},\ and\ \citenamefont {Zerbini}}]{Bertipagani:2020awe}%
  \BibitemOpen
  \bibfield  {author} {\bibinfo {author} {\bibfnamefont {M.}~\bibnamefont {Bertipagani}}, \bibinfo {author} {\bibfnamefont {M.}~\bibnamefont {Rinaldi}}, \bibinfo {author} {\bibfnamefont {L.}~\bibnamefont {Sebastiani}},\ and\ \bibinfo {author} {\bibfnamefont {S.}~\bibnamefont {Zerbini}},\ }\bibfield  {title} {\bibinfo {title} {{Non-singular black holes and mass inflation in modified gravity}},\ }\href {https://doi.org/10.1016/j.dark.2021.100853} {\bibfield  {journal} {\bibinfo  {journal} {Phys. Dark Univ.}\ }\textbf {\bibinfo {volume} {33}},\ \bibinfo {pages} {100853} (\bibinfo {year} {2021})},\ \Eprint {https://arxiv.org/abs/2012.15645} {arXiv:2012.15645 [gr-qc]} \BibitemShut {NoStop}%
\bibitem [{\citenamefont {Bonanno}\ \emph {et~al.}(2021)\citenamefont {Bonanno}, \citenamefont {Khosravi},\ and\ \citenamefont {Saueressig}}]{Bonanno:2020fgp}%
  \BibitemOpen
  \bibfield  {author} {\bibinfo {author} {\bibfnamefont {A.}~\bibnamefont {Bonanno}}, \bibinfo {author} {\bibfnamefont {A.-P.}\ \bibnamefont {Khosravi}},\ and\ \bibinfo {author} {\bibfnamefont {F.}~\bibnamefont {Saueressig}},\ }\bibfield  {title} {\bibinfo {title} {{Regular black holes with stable cores}},\ }\href {https://doi.org/10.1103/PhysRevD.103.124027} {\bibfield  {journal} {\bibinfo  {journal} {Phys. Rev. D}\ }\textbf {\bibinfo {volume} {103}},\ \bibinfo {pages} {124027} (\bibinfo {year} {2021})},\ \Eprint {https://arxiv.org/abs/2010.04226} {arXiv:2010.04226 [gr-qc]} \BibitemShut {NoStop}%
\bibitem [{\citenamefont {Carballo-Rubio}\ \emph {et~al.}(2024)\citenamefont {Carballo-Rubio}, \citenamefont {Di~Filippo}, \citenamefont {Liberati},\ and\ \citenamefont {Visser}}]{Carballo-Rubio:2024dca}%
  \BibitemOpen
  \bibfield  {author} {\bibinfo {author} {\bibfnamefont {R.}~\bibnamefont {Carballo-Rubio}}, \bibinfo {author} {\bibfnamefont {F.}~\bibnamefont {Di~Filippo}}, \bibinfo {author} {\bibfnamefont {S.}~\bibnamefont {Liberati}},\ and\ \bibinfo {author} {\bibfnamefont {M.}~\bibnamefont {Visser}},\ }\bibfield  {title} {\bibinfo {title} {{Mass Inflation without Cauchy Horizons}},\ }\href {https://doi.org/10.1103/PhysRevLett.133.181402} {\bibfield  {journal} {\bibinfo  {journal} {Phys. Rev. Lett.}\ }\textbf {\bibinfo {volume} {133}},\ \bibinfo {pages} {181402} (\bibinfo {year} {2024})},\ \Eprint {https://arxiv.org/abs/2402.14913} {arXiv:2402.14913 [gr-qc]} \BibitemShut {NoStop}%
\bibitem [{\citenamefont {G\"urses}\ \emph {et~al.}(2020)\citenamefont {G\"urses}, \citenamefont {\c{S}i\c{s}man},\ and\ \citenamefont {Tekin}}]{Gurses:2020ofy}%
  \BibitemOpen
  \bibfield  {author} {\bibinfo {author} {\bibfnamefont {M.}~\bibnamefont {G\"urses}}, \bibinfo {author} {\bibfnamefont {T.~c.}\ \bibnamefont {\c{S}i\c{s}man}},\ and\ \bibinfo {author} {\bibfnamefont {B.}~\bibnamefont {Tekin}},\ }\bibfield  {title} {\bibinfo {title} {{Is there a novel Einstein\textendash{}Gauss\textendash{}Bonnet theory in four dimensions?}},\ }\href {https://doi.org/10.1140/epjc/s10052-020-8200-7} {\bibfield  {journal} {\bibinfo  {journal} {Eur. Phys. J. C}\ }\textbf {\bibinfo {volume} {80}},\ \bibinfo {pages} {647} (\bibinfo {year} {2020})},\ \Eprint {https://arxiv.org/abs/2004.03390} {arXiv:2004.03390 [gr-qc]} \BibitemShut {NoStop}%
\bibitem [{\citenamefont {Coll\'eaux}(2020)}]{Colleaux:2020wfv}%
  \BibitemOpen
  \bibfield  {author} {\bibinfo {author} {\bibfnamefont {A.}~\bibnamefont {Coll\'eaux}},\ }\href@noop {} {\bibinfo {title} {{Dimensional aspects of Lovelock-Lanczos gravity}}} (\bibinfo {year} {2020}),\ \Eprint {https://arxiv.org/abs/2010.14174} {arXiv:2010.14174 [gr-qc]} \BibitemShut {NoStop}%
\bibitem [{\citenamefont {Beltran~Jimenez}\ \emph {et~al.}(2016)\citenamefont {Beltran~Jimenez}, \citenamefont {Heisenberg},\ and\ \citenamefont {Koivisto}}]{BeltranJimenez:2016wxw}%
  \BibitemOpen
  \bibfield  {author} {\bibinfo {author} {\bibfnamefont {J.}~\bibnamefont {Beltran~Jimenez}}, \bibinfo {author} {\bibfnamefont {L.}~\bibnamefont {Heisenberg}},\ and\ \bibinfo {author} {\bibfnamefont {T.~S.}\ \bibnamefont {Koivisto}},\ }\bibfield  {title} {\bibinfo {title} {{Cosmology for quadratic gravity in generalized Weyl geometry}},\ }\href {https://doi.org/10.1088/1475-7516/2016/04/046} {\bibfield  {journal} {\bibinfo  {journal} {JCAP}\ }\textbf {\bibinfo {volume} {04}},\ \bibinfo {pages} {046}},\ \Eprint {https://arxiv.org/abs/1602.07287} {arXiv:1602.07287 [hep-th]} \BibitemShut {NoStop}%
\bibitem [{\citenamefont {Beltran~Jimenez}\ and\ \citenamefont {Koivisto}(2014)}]{BeltranJimenez:2014iie}%
  \BibitemOpen
  \bibfield  {author} {\bibinfo {author} {\bibfnamefont {J.}~\bibnamefont {Beltran~Jimenez}}\ and\ \bibinfo {author} {\bibfnamefont {T.~S.}\ \bibnamefont {Koivisto}},\ }\bibfield  {title} {\bibinfo {title} {{Extended Gauss-Bonnet gravities in Weyl geometry}},\ }\href {https://doi.org/10.1088/0264-9381/31/13/135002} {\bibfield  {journal} {\bibinfo  {journal} {Class. Quant. Grav.}\ }\textbf {\bibinfo {volume} {31}},\ \bibinfo {pages} {135002} (\bibinfo {year} {2014})},\ \Eprint {https://arxiv.org/abs/1402.1846} {arXiv:1402.1846 [gr-qc]} \BibitemShut {NoStop}%
\bibitem [{\citenamefont {Bahamonde}\ and\ \citenamefont {Ba{\~n}ados}(2025)}]{Bahamonde:2025qtc}%
  \BibitemOpen
  \bibfield  {author} {\bibinfo {author} {\bibfnamefont {S.}~\bibnamefont {Bahamonde}}\ and\ \bibinfo {author} {\bibfnamefont {M.}~\bibnamefont {Ba{\~n}ados}},\ }\bibfield  {title} {\bibinfo {title} {{An exact five dimensional Weyl-geometry Gauss-Bonnet black hole}},\ }\href {https://doi.org/10.1016/j.physletb.2025.139869} {\bibfield  {journal} {\bibinfo  {journal} {Phys. Lett. B}\ }\textbf {\bibinfo {volume} {869}},\ \bibinfo {pages} {139869} (\bibinfo {year} {2025})},\ \Eprint {https://arxiv.org/abs/2504.02230} {arXiv:2504.02230 [gr-qc]} \BibitemShut {NoStop}%
\bibitem [{\citenamefont {De~Felice}\ \emph {et~al.}(2016{\natexlab{a}})\citenamefont {De~Felice}, \citenamefont {Heisenberg}, \citenamefont {Kase}, \citenamefont {Mukohyama}, \citenamefont {Tsujikawa},\ and\ \citenamefont {Zhang}}]{DeFelice:2016yws}%
  \BibitemOpen
  \bibfield  {author} {\bibinfo {author} {\bibfnamefont {A.}~\bibnamefont {De~Felice}}, \bibinfo {author} {\bibfnamefont {L.}~\bibnamefont {Heisenberg}}, \bibinfo {author} {\bibfnamefont {R.}~\bibnamefont {Kase}}, \bibinfo {author} {\bibfnamefont {S.}~\bibnamefont {Mukohyama}}, \bibinfo {author} {\bibfnamefont {S.}~\bibnamefont {Tsujikawa}},\ and\ \bibinfo {author} {\bibfnamefont {Y.-l.}\ \bibnamefont {Zhang}},\ }\bibfield  {title} {\bibinfo {title} {{Cosmology in generalized Proca theories}},\ }\href {https://doi.org/10.1088/1475-7516/2016/06/048} {\bibfield  {journal} {\bibinfo  {journal} {JCAP}\ }\textbf {\bibinfo {volume} {06}},\ \bibinfo {pages} {048}},\ \Eprint {https://arxiv.org/abs/1603.05806} {arXiv:1603.05806 [gr-qc]} \BibitemShut {NoStop}%
\bibitem [{\citenamefont {De~Felice}\ \emph {et~al.}(2016{\natexlab{b}})\citenamefont {De~Felice}, \citenamefont {Heisenberg}, \citenamefont {Kase}, \citenamefont {Mukohyama}, \citenamefont {Tsujikawa},\ and\ \citenamefont {Zhang}}]{DeFelice:2016uil}%
  \BibitemOpen
  \bibfield  {author} {\bibinfo {author} {\bibfnamefont {A.}~\bibnamefont {De~Felice}}, \bibinfo {author} {\bibfnamefont {L.}~\bibnamefont {Heisenberg}}, \bibinfo {author} {\bibfnamefont {R.}~\bibnamefont {Kase}}, \bibinfo {author} {\bibfnamefont {S.}~\bibnamefont {Mukohyama}}, \bibinfo {author} {\bibfnamefont {S.}~\bibnamefont {Tsujikawa}},\ and\ \bibinfo {author} {\bibfnamefont {Y.-l.}\ \bibnamefont {Zhang}},\ }\bibfield  {title} {\bibinfo {title} {{Effective gravitational couplings for cosmological perturbations in generalized Proca theories}},\ }\href {https://doi.org/10.1103/PhysRevD.94.044024} {\bibfield  {journal} {\bibinfo  {journal} {Phys. Rev. D}\ }\textbf {\bibinfo {volume} {94}},\ \bibinfo {pages} {044024} (\bibinfo {year} {2016}{\natexlab{b}})},\ \Eprint {https://arxiv.org/abs/1605.05066} {arXiv:1605.05066 [gr-qc]} \BibitemShut {NoStop}%
\bibitem [{\citenamefont {de~Felice}\ \emph {et~al.}(2017)\citenamefont {de~Felice}, \citenamefont {Heisenberg},\ and\ \citenamefont {Tsujikawa}}]{deFelice:2017paw}%
  \BibitemOpen
  \bibfield  {author} {\bibinfo {author} {\bibfnamefont {A.}~\bibnamefont {de~Felice}}, \bibinfo {author} {\bibfnamefont {L.}~\bibnamefont {Heisenberg}},\ and\ \bibinfo {author} {\bibfnamefont {S.}~\bibnamefont {Tsujikawa}},\ }\bibfield  {title} {\bibinfo {title} {{Observational constraints on generalized Proca theories}},\ }\href {https://doi.org/10.1103/PhysRevD.95.123540} {\bibfield  {journal} {\bibinfo  {journal} {Phys. Rev. D}\ }\textbf {\bibinfo {volume} {95}},\ \bibinfo {pages} {123540} (\bibinfo {year} {2017})},\ \Eprint {https://arxiv.org/abs/1703.09573} {arXiv:1703.09573 [astro-ph.CO]} \BibitemShut {NoStop}%
\bibitem [{\citenamefont {Heisenberg}\ and\ \citenamefont {Villarrubia-Rojo}(2021)}]{Heisenberg:2020xak}%
  \BibitemOpen
  \bibfield  {author} {\bibinfo {author} {\bibfnamefont {L.}~\bibnamefont {Heisenberg}}\ and\ \bibinfo {author} {\bibfnamefont {H.}~\bibnamefont {Villarrubia-Rojo}},\ }\bibfield  {title} {\bibinfo {title} {{Proca in the sky}},\ }\href {https://doi.org/10.1088/1475-7516/2021/03/032} {\bibfield  {journal} {\bibinfo  {journal} {JCAP}\ }\textbf {\bibinfo {volume} {03}},\ \bibinfo {pages} {032}},\ \Eprint {https://arxiv.org/abs/2010.00513} {arXiv:2010.00513 [astro-ph.CO]} \BibitemShut {NoStop}%
\bibitem [{\citenamefont {Bohnenblust}\ \emph {et~al.}(2025)\citenamefont {Bohnenblust}, \citenamefont {Giardino}, \citenamefont {Heisenberg},\ and\ \citenamefont {Nussbaumer}}]{Bohnenblust:2024mou}%
  \BibitemOpen
  \bibfield  {author} {\bibinfo {author} {\bibfnamefont {L.}~\bibnamefont {Bohnenblust}}, \bibinfo {author} {\bibfnamefont {S.}~\bibnamefont {Giardino}}, \bibinfo {author} {\bibfnamefont {L.}~\bibnamefont {Heisenberg}},\ and\ \bibinfo {author} {\bibfnamefont {N.}~\bibnamefont {Nussbaumer}},\ }\bibfield  {title} {\bibinfo {title} {{To bounce or not to bounce in generalized Proca theory and beyond}},\ }\href {https://doi.org/10.1007/JHEP07(2025)124} {\bibfield  {journal} {\bibinfo  {journal} {JHEP}\ }\textbf {\bibinfo {volume} {07}},\ \bibinfo {pages} {124}},\ \Eprint {https://arxiv.org/abs/2412.03977} {arXiv:2412.03977 [gr-qc]} \BibitemShut {NoStop}%
\bibitem [{\citenamefont {Birmingham}(1999)}]{Birmingham_1999}%
  \BibitemOpen
  \bibfield  {author} {\bibinfo {author} {\bibfnamefont {D.}~\bibnamefont {Birmingham}},\ }\bibfield  {title} {\bibinfo {title} {Topological black holes in anti-de sitter space},\ }\href {https://doi.org/10.1088/0264-9381/16/4/009} {\bibfield  {journal} {\bibinfo  {journal} {Classical and Quantum Gravity}\ }\textbf {\bibinfo {volume} {16}},\ \bibinfo {pages} {1197–1205} (\bibinfo {year} {1999})}\BibitemShut {NoStop}%
\bibitem [{\citenamefont {Heisenberg}\ \emph {et~al.}(2017{\natexlab{a}})\citenamefont {Heisenberg}, \citenamefont {Kase}, \citenamefont {Minamitsuji},\ and\ \citenamefont {Tsujikawa}}]{Heisenberg:2017hwb}%
  \BibitemOpen
  \bibfield  {author} {\bibinfo {author} {\bibfnamefont {L.}~\bibnamefont {Heisenberg}}, \bibinfo {author} {\bibfnamefont {R.}~\bibnamefont {Kase}}, \bibinfo {author} {\bibfnamefont {M.}~\bibnamefont {Minamitsuji}},\ and\ \bibinfo {author} {\bibfnamefont {S.}~\bibnamefont {Tsujikawa}},\ }\bibfield  {title} {\bibinfo {title} {{Black holes in vector-tensor theories}},\ }\href {https://doi.org/10.1088/1475-7516/2017/08/024} {\bibfield  {journal} {\bibinfo  {journal} {JCAP}\ }\textbf {\bibinfo {volume} {08}},\ \bibinfo {pages} {024}},\ \Eprint {https://arxiv.org/abs/1706.05115} {arXiv:1706.05115 [gr-qc]} \BibitemShut {NoStop}%
\bibitem [{\citenamefont {Heisenberg}\ \emph {et~al.}(2017{\natexlab{b}})\citenamefont {Heisenberg}, \citenamefont {Kase}, \citenamefont {Minamitsuji},\ and\ \citenamefont {Tsujikawa}}]{Heisenberg:2017xda}%
  \BibitemOpen
  \bibfield  {author} {\bibinfo {author} {\bibfnamefont {L.}~\bibnamefont {Heisenberg}}, \bibinfo {author} {\bibfnamefont {R.}~\bibnamefont {Kase}}, \bibinfo {author} {\bibfnamefont {M.}~\bibnamefont {Minamitsuji}},\ and\ \bibinfo {author} {\bibfnamefont {S.}~\bibnamefont {Tsujikawa}},\ }\bibfield  {title} {\bibinfo {title} {{Hairy black-hole solutions in generalized Proca theories}},\ }\href {https://doi.org/10.1103/PhysRevD.96.084049} {\bibfield  {journal} {\bibinfo  {journal} {Phys. Rev. D}\ }\textbf {\bibinfo {volume} {96}},\ \bibinfo {pages} {084049} (\bibinfo {year} {2017}{\natexlab{b}})},\ \Eprint {https://arxiv.org/abs/1705.09662} {arXiv:1705.09662 [gr-qc]} \BibitemShut {NoStop}%
\bibitem [{\citenamefont {De~Felice}\ \emph {et~al.}(2016{\natexlab{c}})\citenamefont {De~Felice}, \citenamefont {Heisenberg}, \citenamefont {Kase}, \citenamefont {Tsujikawa}, \citenamefont {Zhang},\ and\ \citenamefont {Zhao}}]{DeFelice:2016cri}%
  \BibitemOpen
  \bibfield  {author} {\bibinfo {author} {\bibfnamefont {A.}~\bibnamefont {De~Felice}}, \bibinfo {author} {\bibfnamefont {L.}~\bibnamefont {Heisenberg}}, \bibinfo {author} {\bibfnamefont {R.}~\bibnamefont {Kase}}, \bibinfo {author} {\bibfnamefont {S.}~\bibnamefont {Tsujikawa}}, \bibinfo {author} {\bibfnamefont {Y.-l.}\ \bibnamefont {Zhang}},\ and\ \bibinfo {author} {\bibfnamefont {G.-B.}\ \bibnamefont {Zhao}},\ }\bibfield  {title} {\bibinfo {title} {{Screening fifth forces in generalized Proca theories}},\ }\href {https://doi.org/10.1103/PhysRevD.93.104016} {\bibfield  {journal} {\bibinfo  {journal} {Phys. Rev. D}\ }\textbf {\bibinfo {volume} {93}},\ \bibinfo {pages} {104016} (\bibinfo {year} {2016}{\natexlab{c}})},\ \Eprint {https://arxiv.org/abs/1602.00371} {arXiv:1602.00371 [gr-qc]} \BibitemShut {NoStop}%
\bibitem [{\citenamefont {Fazzini}\ \emph {et~al.}(2023)\citenamefont {Fazzini}, \citenamefont {Rovelli},\ and\ \citenamefont {Soltani}}]{Fazzini:2023scu}%
  \BibitemOpen
  \bibfield  {author} {\bibinfo {author} {\bibfnamefont {F.}~\bibnamefont {Fazzini}}, \bibinfo {author} {\bibfnamefont {C.}~\bibnamefont {Rovelli}},\ and\ \bibinfo {author} {\bibfnamefont {F.}~\bibnamefont {Soltani}},\ }\bibfield  {title} {\bibinfo {title} {{Painlev{\'e}-Gullstrand coordinates discontinuity in the quantum Oppenheimer-Snyder model}},\ }\href {https://doi.org/10.1103/PhysRevD.108.044009} {\bibfield  {journal} {\bibinfo  {journal} {Phys. Rev. D}\ }\textbf {\bibinfo {volume} {108}},\ \bibinfo {pages} {044009} (\bibinfo {year} {2023})},\ \Eprint {https://arxiv.org/abs/2307.07797} {arXiv:2307.07797 [gr-qc]} \BibitemShut {NoStop}%
\bibitem [{\citenamefont {Rovelli}\ and\ \citenamefont {Vidotto}(2024)}]{Rovelli:2024sjl}%
  \BibitemOpen
  \bibfield  {author} {\bibinfo {author} {\bibfnamefont {C.}~\bibnamefont {Rovelli}}\ and\ \bibinfo {author} {\bibfnamefont {F.}~\bibnamefont {Vidotto}},\ }\href@noop {} {\bibinfo {title} {{Planck stars, White Holes, Remnants and Planck-mass quasi-particles. The quantum gravity phase in black holes' evolution and its manifestations}}} (\bibinfo {year} {2024}),\ \Eprint {https://arxiv.org/abs/2407.09584} {arXiv:2407.09584 [gr-qc]} \BibitemShut {NoStop}%
\end{thebibliography}%

\end{document}